\documentclass[11pt,aps,nofootinbib,floatfix,pra,tightenlines]{revtex4}%

\usepackage{amsmath}
\usepackage{amsfonts}
\usepackage{amssymb}
\usepackage{graphicx}
\usepackage{graphics}
\usepackage{epsfig}
\usepackage[section]{placeins}
\usepackage{multirow}
\usepackage{bibentry}
\usepackage{braket}
\usepackage[font=footnotesize,labelfont=bf]{caption}
\usepackage{pdfpages}
\usepackage{slashed}
\usepackage{bbm} 
\providecommand{\U}[1]{\protect\rule{.1in}{.1in}}
\nobibliography*

\newcommand{\im}{\text{i}}

\newcommand{\eref}[1]{Eq.~(\ref{#1})}

\def\){\right)} 
\def\({\left(} 
\def\]{\right]} 
\def\[{\left[}

\DeclareMathOperator{\Tr}{Tr}

\begin{document}

\title{Alpha-alpha scattering in the Multiverse}

\author{Serdar Elhatisari}
\email{selhatisari@gmail.com}
\affiliation{Faculty of Natural Sciences and Engineering, Gaziantep Islam Science and
Technology University, Gaziantep 27010, Turkey}

\author{Timo A. L\"ahde}
\email{t.laehde@fz-juelich.de}
\affiliation{
Institute~for~Advanced~Simulation, Institut~f\"{u}r~Kernphysik, 
Center for Advanced Simulation and Analytics, and
J\"{u}lich~Center~for~Hadron~Physics, \\ 
Forschungszentrum~J\"{u}lich, D-52425~J\"{u}lich,~Germany}

\author{Dean Lee}
\email{leed@frib.msu.edu}
\affiliation{Facility for Rare Isotope Beams and Department of Physics and Astronomy,
Michigan State University, East Lansing, MI 48824, USA}

\author{Ulf-G. Mei{\ss}ner}
\email{meissner@hiskp.uni-bonn.de}
\affiliation{Helmholtz-Institut f{\"u}r Strahlen- und Kernphysik and Bethe Center
for Theoretical Physics, Universit\"at Bonn, D-53115 Bonn, Germany\\
Institute~for~Advanced~Simulation, Institut~f\"{u}r~Kernphysik,
Center for Advanced Simulation and Analytics, and
J\"{u}lich~Center~for~Hadron~Physics, \\ 
Forschungszentrum~J\"{u}lich, D-52425~J\"{u}lich,~Germany\\
Tbilisi State University, 0186 Tbilisi, Georgia}

\author{Thomas Vonk}
\email{vonk@hiskp.uni-bonn.de}
\affiliation{Helmholtz-Institut f{\"u}r Strahlen- und Kernphysik and Bethe Center
for Theoretical Physics, Universit\"at Bonn, D-53115 Bonn, Germany}

\begin{abstract}
We investigate the phase shifts of low-energy $\alpha$-$\alpha$ scattering under variations
of the fundamental parameters of the Standard Model, namely the light quark mass, the electromagnetic
fine-structure constant as well as the QCD $\theta$-angle. As a first step, we recalculate
$\alpha$-$\alpha$ scattering in our Universe utilizing various improvements in the adiabatic
projection method, which leads to an improved, parameter-free prediction of the S- and D-wave
phase shifts for laboratory energies below 10~MeV. We find that positive shifts in the pion mass
have a small effect on the S-wave phase shift, whereas lowering the pion mass adds some
repulsion in the two-alpha system. The effect on the D-wave phase shift turns
out to be more pronounced as signaled by the D-wave resonance parameters. Variations of the fine-structure
constant have almost no effect on the low-energy $\alpha$-$\alpha$ phase shifts. We further show that
up-to-and-including next-to-leading order in the chiral expansion, variations of these phase shifts
with respect to the QCD $\theta$-angle can be expressed in terms of the $\theta$-dependent pion mass.
\end{abstract}

\date{\today}

\maketitle


\section{Introduction}
\label{sec:intro}

Alpha-alpha ($\alpha$-$\alpha$) scattering is one of the most fundamental reactions in nuclear (astro)physics.
It is the basic component of the triple-alpha (3$\alpha$) reaction prevalent in hot old stars, that leads
to the generation of $^{12}$C and successively $^{16}$O, where the $^{12}$C production is enhanced through
a $J^P=0^+$ resonance at $7.65$~MeV excitation energy close to the $3\alpha$-threshold, the famous
Hoyle state \cite{Hoyle:1954zz}. $\alpha$-$\alpha$ scattering itself features some fine-tuning, as the
large near-threshold S-wave results from a state with $(J^P,I) = (0^+,0)$ at an energy $E_R\simeq 0.1$~MeV
above the threshold, see e.g. the review \cite{Afzal:1969hal},
with a tiny width of $\Gamma_R\simeq 6$~eV. It is precisely this small width
(long lifetime) of the unstable $^8$Be nucleus that allows for the reaction with the third $\alpha$ particle
in the $3\alpha$ reaction at sufficiently high temperatures and densities.

The fine-tunings in these (and other) fundamental nuclear reactions together with other fine-tunings in
particle physics and cosmology have led to the concept of the {\em Multiverse}, where our Universe with
its observed values is part of a larger structure of universes featuring different sets of the fundamental
constants.
Related to this are anthropic considerations, which is the philosophical idea that the parameters governing
our world should fit the intervals compatible with the existence of life on Earth. More details can
be found in the reviews~\cite{Hogan:1999wh,Schellekens:2013bpa,Meissner:2014pma,Donoghue:2016tjk,Adams:2019kby}.

Coming back to nuclear physics, the closeness of the Hoyle state energy to the $3\alpha$ threshold
invites investigations about the stability of this resonance condition under changes of the fundamental
parameters of the strong and the electromagnetic (EM) interactions, whose interplay guarantees the stability
of atomic nuclei. While earlier investigations, see e.g. Ref.~\cite{Oberhummer:2000zj}, suffered from
some model-dependence in the description of the nuclear forces, using the {\em ab initio} method of
Nuclear Lattice Effective Field Theory (NLEFT) this topic was re-investigated in
Refs.~\cite{Epelbaum:2012iu,Epelbaum:2013wla,Lahde:2019yvr}. More specifically, the quark mass dependence
as well as the
dependence on the electromagnetic fine-structure constant of the nuclear Hamiltonian was worked out,
using and combining results from chiral perturbation theory (CHPT) and lattice QCD simulations for the pion decay
constant, the nucleon mass and so on. Here, we will use the same chiral EFT at next-to-next-to-leading
order combined with the so-called Adiabatic Projection Method (APM), that allows for  {\em ab initio}
calculations of nuclear reactions, as developed in Refs.~\cite{Pine:2013zja,Elhatisari:2014lka,Rokash:2015hra}.
Using the APM, the scattering of two alpha clusters has been achieved on the lattice \cite{Elhatisari:2015iga},
enabled by the fact that the computational effort is approximately quadratic in the number of nucleons in the
scattering clusters. The method was further refined in Ref.~\cite{Elhatisari:2016hby}. Combining these
different works,
we are thus in the position to investigate the sensitivity of the low-energy $\alpha$-$\alpha$ phase shifts
on variations in the light quark mass $\hat{m}$ and the em fine-structure constant $\alpha_{\rm EM}$.  We
note that $\alpha$-$\alpha$ scattering has also recently been studied using the no-core shell model
within a continuum approach \cite{Kravvaris:2020cvn}.

While the investigation of the resonance enhancement in the $3\alpha$ process due to the Hoyle
state already sets rather stringent limits on the possible variations of the light quark mass and
the fine-structure constant, one has to be aware that these results are afflicted with some inherent
uncertainties, as in the corresponding stellar simulations only the distance of the Hoyle state
to the $3\alpha$-threshold is varied. Translating this into a dependence on, say, the light quark mass
assumes that only the nuclei directly involved in the $3\alpha$ process are subject to these changes,
but of course one should perform the complete stellar simulations (reaction networks) with appropriately
modified masses and reaction rates. At present, this is only possible for Big Bang Nucleosynthesis, see e.g.
Refs.~\cite{Bedaque:2010hr,Berengut:2013nh}, but not for the whole nuclear reaction networks in stars.
Therefore, the {\em ab initio} computation of the dependence of $\alpha$-$\alpha$ scattering on the
fundamental parameters of the Standard Model is not subject to such uncertainties and paves the
way for more elaborate network calculations in the Multiverse.

A parameter that has obtained less attention in such anthropic considerations is the QCD $\theta$-term,
as the bounds from the neutron electric dipole moment require $\theta \lesssim 10^{-10}$, see e.g.
Ref.~\cite{Dragos:2019oxn} for a recent lattice QCD study. Still, it is worth to reconsider bounds
on the $\theta$-angle from observations other than the neutron EDM as well as from anthropic considerations,
as done e.g. in Refs.~\cite{Ubaldi:2008nf,Lee:2020tmi}. In particular, it was shown in \cite{Lee:2020tmi}
that nuclear binding increases with $\theta$ and that $\theta \lesssim 0.1$ would not upset the world as we
know it. It is thus also of interest to study the reaction rate of the fundamental $\alpha$-$\alpha$ scattering
process as a function of $\theta$, as will be done here.

In Ref.~\cite{Elhatisari:2016owd}, it was shown that symmetric nuclear matter without Coulomb interactions
lies close to a quantum phase transition between a Bose gas of alpha clusters and a nuclear liquid.
Whether one is in the Bose gas phase or the nuclear liquid phase is determined by the sign of the
$\alpha$-$\alpha$ S-wave scattering length.  In turn, the $\alpha$-$\alpha$ scattering phase shifts depend
on the strength, range, and locality of the nucleon-nucleon interactions.  The nucleon-nucleon interactions
need enough attractive strength, range, and locality to overcome the Pauli repulsion between nucleons with
the same spin and isospin  \cite{Rokash:2016tqh,Kanada-Enyo:2020zzf}.  Locality here refers to interactions
that are diagonal when written in position space. The variation of the light quark masses, eletromagnetic
fine-structure constant, and $\theta$ parameter will produce changes to the leading-order interactions,
and we take these changes to the nucleon-nucleon interactions to be local.  This choice is motivated by
studies of Quantum Chromodynamics in the limit of a large number of colors showing that the
nucleon-nucleon interactions reduce to local interactions with an underlying spin-isospin
exchange symmetry \cite{Kaplan:1995yg,Kaplan:1996rk,Lee:2020esp}. 

The paper is organized as follows. In Sec.~\ref{sec:basics}, we introduce the dependence of the
two-alpha cluster energy on the fundamental parameters of the Standard Model, the basic framework of NLEFT
and give a first glimpse on some of the relevant quark (pion) mass dependences. The pion mass
dependence of the nuclear Hamiltonian used here is presented in detail in Sec.~\ref{sec:Mpi-theta-dependence}.
Then, in Sec.~\ref{sec:alpha-em-dependence} we discuss the inclusion of the electromagnetic interaction
and the dependence of the nuclear Hamiltonian on the fine-structure constant. Sec.~\ref{sec:theta} shows
how the $\theta$-dependence of $\alpha$-$\alpha$ scattering can be inferred from the $\theta$-dependence
of the pion mass. In Sec.~\ref{sec:tools} we collect the computational tools needed for this investigations.
We give the basic APM formalism  needed for our investigation and  show how various quantities are obtained
from Auxiliary Field Quantum Monte Carlo simulations. In Sec.~\ref{sec:APM-SPS}, we show how to
extract the scattering phase shifts from the adiabatic transfer matrices.
Our results are presented and discussed in Sec.~\ref{sec:results}.
We end with a summary and conclusions. Some further details of the computations are relegated to the
appendices.

\section{Basic concepts}
\label{sec:basics}

We aim to compute the variation of the $\alpha$-$\alpha$ scattering phase shifts as a
function of the fundamental constants of nature following Refs.~\cite{Epelbaum:2013wla,Lahde:2019yvr}.
Since we compute the scattering phase shifts from the spectrum, we consider a linear variation
in  the light quark mass and the electromagnetic fine-structure constant $\alpha_{\rm EM}$
of the two-alpha cluster energy,
\begin{align}
\delta E_{\alpha\alpha} 
\simeq 
\left. \frac{\partial E_{\alpha\alpha}}
{\partial M_{\pi}}          \right|_{M_{\pi}^{\rm ph}} \delta M_{\pi}
+ \left. \frac{\partial E_{\alpha\alpha}}
{\partial \alpha_{\rm EM}}  \right|_{\alpha_{\rm EM}^{\rm ph}} \delta \alpha_{\rm EM}~,
\label{eqn:0001}%
\end{align}
where we have used the Gell-Mann--Oakes--Renner relation, $M^2_\pi = 2 B_0 \hat{m}$, with $\hat{m}=(m_u+m_d)/2$
the light quark mass and $B_0$ is related to the scalar quark condensate.\footnote{
Because of this relation, we can equivalently use the wordings ``quark mass dependence'' and ``pion mass dependence''.}
Throughout, we work in the isospin limit as strong isospin breaking effects are expected to be very small. 
Further, the superscript ``ph'' denotes the pertinent values in Nature (the physical world). We note that this  formula is applicable
for changes in the modulus of the pion mass $|\delta M_\pi/M_\pi|$ and the electromagnetic fine-structure constant
by $|\delta\alpha_{\rm EM}/\alpha_{\rm EM}|\lesssim 10\%$. The variation
with respect to the QCD $\theta$ angle will be discussed later in a separate section.

Our computational framework is NLEFT, see Refs.~\cite{Lee:2008fa,Lahde:2019npb} for details.
In what follows, we employ a periodic cubic lattice with a spatial lattice spacing of $a = 1.97$~fm
and a temporal lattice spacing $a_t = 1.32$~fm. 
For free nucleons we use the $\mathcal{O}(a^4)$-improved lattice Hamiltonian, 
\begin{eqnarray}
H_{\rm free} &= 
&\frac{49}{12 m_{N}}
\sum_{\vec{n}}\sum_{i,j=0,1} a_{i,j}^{\dagger}(\vec{n}) a_{i,j}^{\,}(\vec{n})
\nonumber\\
&
-&\frac{3}{4 m_{N}}
\sum_{\vec{n}}\sum_{i,j=0,1}\sum_{l=1,2,3}
\left[a_{i,j}^{\dagger}(\vec{n}) a_{i,j}^{\,}(\vec{n}+\hat{l})
+a_{i,j}^{\dagger}(\vec{n}) a_{i,j}^{\,}(\vec{n}\hat{l})\right]
\nonumber\\
&
-&\frac{3}{40 m_{N}}
\sum_{\vec{n}}\sum_{i,j=0,1}\sum_{l=1,2,3}
\left[a_{i,j}^{\dagger}(\vec{n}) a_{i,j}^{\,}(\vec{n}+2\hat{l})
+a_{i,j}^{\dagger}(\vec{n}) a_{i,j}^{\,}(\vec{n}-2\hat{l})\right]
\nonumber
\end{eqnarray}
\begin{eqnarray}
&
-&\frac{1}{180 m_{N}}
\sum_{\vec{n}}\sum_{i,j=0,1}\sum_{l=1,2,3}
\left[a_{i,j}^{\dagger}(\vec{n}) a_{i,j}^{\,}(\vec{n}+3\hat{l})
+a_{i,j}^{\dagger}(\vec{n}) a_{i,j}^{\,}(\vec{n}-3\hat{l})\right]~,
\label{eqn:0001a}%
\end{eqnarray}
where $\vec{n}$ represents the integer-valued lattice sites, $m_N$ is the nucleon mass,
$\hat{l} = \hat{1}, \hat{2}, \hat{3}$ are unit lattice vectors in the spatial directions,
$i(j)$ is a spin (isospin) index, and $a_{i,j}^{\,}$ and $a_{i,j}^{\dagger}$ denote nucleon
annihilation and creation operators. 

For the leading-order (LO) nuclear interaction we use an improved action which is based on the
following nucleon-nucleon (NN) scattering amplitude,
\begin{align}
\mathcal{A}_{\rm LO}
= & 
C_{S=0,I=1} \, f(\vec{q}) \, \left(\frac{1}{4}-\frac{1}{4}\vec{\sigma}_i\cdot\vec{\sigma}_j\right)
\left(\frac{3}{4}+\frac{1}{4}\vec{\tau}_i\cdot\vec{\tau}_j\right)
\nonumber\\
& + C_{S=1,I=0} \, f(\vec{q}) \, \left(\frac{3}{4}+\frac{1}{4}\vec{\sigma}_i\cdot\vec{\sigma}_j\right)
\left(\frac{1}{4}-\frac{1}{4}\vec{\tau}_i\cdot\vec{\tau}_j\right)
\nonumber\\
& + \tilde{g}_{\pi N}^2 \vec{\tau}_i\cdot\vec{\tau}_j \frac{(\vec{\sigma}_i\cdot \vec{q})(\vec{\sigma}_j
  \cdot \vec{q})} {\vec{q}^2 + M_{\pi}^2}
\,,
\label{eqn:0001b}%
\end{align}
where $\vec{\sigma}$ and $\vec{\tau}$ denote the Pauli spin and isospin matrices, $\tilde{g}_{\pi N}$
is the strength of the one-pion-exchange (OPE) potential defined as $\tilde{g}_{\pi N}
={g_{A}}/(2\, F_{\pi})$ in terms of the nucleon axial-vector coupling $g_A =1.273(19)$ and the pion
decay constant $F_{\pi}= 92.1$~MeV. 
$C_{S=0,I=1}$ and $C_{S=1,I=0}$ are the coupling constants of the short-range  part of the nuclear force which
are adjusted to reproduce the scattering phase shifts for the two S-wave channels, and
$f(\vec{q})$ is a smearing function which is defined to reproduce the effective ranges for  the
two S-wave channels. We redefine the low-energy constants (LECs) of the short-range interactions in terms
of linear combinations of $C_0$ and $C_{I}$, 
\begin{align}
C_{0} & = \frac{3}{4} C_{S=0,I=1} + \frac{1}{4} C_{S=1,I=0}\,,
\\
C_{I} & = \frac{1}{4} C_{S=0,I=1} - \frac{3}{4} C_{S=1,I=0}\,.
\label{eqn:0001c}%
\end{align}
From Eqs.~(\ref{eqn:0001a}) and (\ref{eqn:0001b}) it is obvious that the sources of implicit $M_{\pi}$-dependence
are the nucleon mass $m_{N}$, the coupling constant of the OPE potential
$\tilde{g}_{\pi N}$, and the LECs of the short-range interactions $C_0$ and $C_{I}$, besides the explicit
pion mass dependence in the OPE.
Before discussing these in detail in Sec.~\ref{sec:Mpi-theta-dependence}, let us consider
the quark (pion) mass dependence of the nucleon mass and the pion decay constant to get an
idea about the changes we can expect. At the leading one-loop order ${\cal O}(p^3)$, where $p$
is a generic small parameter, the chiral expansion of the nucleon mass can be written as
\begin{align}
m_{N}(M_{\pi}) = m_{0} - 4 c_{1} M_{\pi}^{2} -\frac{3 g_{A}^{2}(M_\pi) M_{\pi}^{3}}{32 \pi F_{\pi}^{2}(M_{\pi})}
+ \mathcal{O}(M_{\pi}^4)\,,
\label{eqn:0001d}%
\end{align}
where $m_{0}\simeq 865$~MeV~\cite{Hoferichter:2015hva} is the nucleon mass in the
(two-flavor) chiral limit and $c_{1}=-1.1$~GeV$^{-1}$ is a LEC from the chiral pion-nucleon Lagrangian at
next-to-leading order (NLO)~\cite{Hoferichter:2015tha}.
Note that the leading correction of order $M_\pi^2$ is intimately linked to the pion-nucleon
$\sigma$-term discussed below. At third order, the pion mass dependence of the pion decay constant
and the axial-vector coupling constant is made explicit.
For the pion decay constant we use the expression from the chiral
expansion at NLO, 
\begin{align}
F_{\pi}(M_{\pi}) = F +  \frac{M_{\pi}^{2}}{16 \pi^2 F} \bar{l}_{4} + \mathcal{O}(M_{\pi}^4)\,,
\label{eqn:0001e}%
\end{align}
where $F=86.2$~MeV is the pion decay constant in the (two-flavor) chiral limit,\footnote{Note
that throughout we do not consider variations of the strange quark mass $m_s$, as these are expected to
  be very small. Hence $m_s$ is simply kept at its physical value.}
and $\bar{l}_4=4.3$ is a LEC, where we use the value from Ref.~\cite{Gasser:1983yg}
(which is consistent with more modern determinations).
We postpone the discussion of the nucleon axial-vector coupling $g_{A}$ and of the LECs $C_0, C_I$ to
the next section.

\section{Pion mass dependence of the nuclear Hamiltonian}
\label{sec:Mpi-theta-dependence}

First, let us collect the knowledge about the pion mass dependence of the nuclear Hamiltonian.
Specifically, the dependence of the energy $E_{\alpha\alpha}$ on the pion mass $M_{\pi}$ can be expressed as 
\begin{align}
E_{\alpha\alpha} 
=
E_{\alpha\alpha}(\tilde{M}_{\pi}, m_{N}({M}_{\pi}), \tilde{g}_{\pi N}({M}_{\pi}), C_{0}({M}_{\pi}), C_{I}({M}_{\pi})),
\label{eqn:0005}%
\end{align}
where $\tilde{M}_{\pi}$ denotes the explicit ${M}_{\pi}$-dependence from the pion propagator in the
OPE potential. Without going into the details of the individual terms
given here, we write the variation of the two-alpha cluster energy around the physical point as
\begin{align}
\left. \frac{\partial E_{\alpha\alpha}}
{\partial M_{\pi}}          \right|_{M_{\pi}^{\rm ph}}
=
&
\left. \frac{\partial E_{\alpha\alpha}}
{\partial \tilde{M}_{\pi}}          \right|_{M_{\pi}^{\rm ph}}
+
x_{1} \,
\left. \frac{\partial E_{\alpha\alpha}}
{\partial m_{N}}          \right|_{m_{N}^{\rm ph}}
\nonumber\\
&
+
x_{2} \,
\left. \frac{\partial E_{\alpha\alpha}}
{\partial \tilde{g}_{\pi N}}          \right|_{\tilde{g}_{\pi N}^{\rm ph}}
+ 
x_{3} \,
\left. \frac{\partial E_{\alpha\alpha}}
{\partial C_{0}}          \right|_{C_{0}^{\rm ph}}
+
x_{4} \,
\left. \frac{\partial E_{\alpha\alpha}}
{\partial C_{I}}          \right|_{C_{I}^{\rm ph}}\,,
\label{eqn:0009}%
\end{align}
where
\begin{align}
&x_{1} = 
\left. \frac{\partial m_{N}}
{\partial M_{\pi}} \right|_{M_{\pi}^{\rm ph}}\, , \,
x_{2} = 
\left. \frac{\partial \tilde{g}_{\pi N}}
{\partial M_{\pi}} \right|_{M_{\pi}^{\rm ph}}\, , \,
x_{3} = 
\left. \frac{\partial C_0}
{\partial M_{\pi}} \right|_{M_{\pi}^{\rm ph}}\, , \,
x_{4} = 
\left. \frac{\partial C_I}
{\partial M_{\pi}} \right|_{M_{\pi}^{\rm ph}}\, . \,
\label{eqn:0009a}
\end{align}

The partial derivatives in Eq.~(\ref{eqn:0009}) are computed using the auxiliary field quantum Monte
Carlo (AFQMC) method~\cite{Lee:2008fa}, see Sec.~\ref{sec:tools}. To obtain an accurate and
model-independent description of the $M_{\pi}$-dependence of the LO nuclear interaction,
we will use the most recent knowledge from chiral perturbation theory and lattice QCD simulations to
determine the quantities in Eq.~(\ref{eqn:0009a}).

The partial derivative $\partial E_{\alpha\alpha} / \partial \tilde{M}_{\pi} $ in Eq.~(\ref{eqn:0009})
is computed by introducing a small change in the pion mass in the OPE of the nuclear Hamiltonian,
$H(\tilde{M}_{\pi}) \to H(\tilde{M}_{\pi} + \Delta \tilde{M}_{\pi})$, which corresponds to a perturbative
shift in the energy, $\Delta E_{\alpha\alpha}(\tilde{M}_{\pi})$. In our calculations, the pion masses are
shifted by $\Delta \tilde{M}_{\pi}=4.59$~MeV, which equals to the empirical mass difference between the
neutral and charged pions. Therefore, the partial derivative $\partial E_{\alpha\alpha} / \partial
\tilde{M}_{\pi}$ is defined as
\begin{align}
\left. \frac{\partial E_{\alpha\alpha}}
{\partial \tilde{M}_{\pi}}          \right|_{M_{\pi}^{\rm ph}}
=
\frac{\Delta E_{\alpha\alpha}(\tilde{M}_{\pi})}
{\Delta \tilde{M}_{\pi}}.
\label{eqn:0013}%
\end{align}

\noindent
In what follows, we will also use the so-called $K$-factors.  These are defined via
\begin{equation}
K_X^i = \left.\frac{y}{X}\frac{\partial X}{\partial y}\right|_{y^{\rm ph}}~,
\end{equation}  
where $X$ is an observable and the superscript $i = \{q, \pi, \alpha\}$ denotes the quantity
$y=\{m_q , M_\pi, \alpha_{\rm EM}\}$, such that, e.g., $K_X^q$ measures the sensitivity of $X$ to changes in
the light quark mass $m_q$. For more detailed discussion on these quantities, see, e.g.,
Ref.~\cite{Berengut:2013nh}.

The parameter $x_{1}$ can be determined from the pion-nucleon sigma term,
\begin{equation}
\sigma_{\pi N}= \langle N|\hat{m}(\bar{u}u+\bar{d}d|N\rangle =
M_{\pi}^2\, \frac{\partial m_{N}}{\partial M^2_{\pi}}~,
\end{equation}  
i.e. the quark mass dependence of the nucleon mass, via
\begin{align}
x_{1}  = \left. \frac{\partial m_{N}}
{\partial M_{\pi}} \right|_{M_{\pi}^{\rm ph}} 
= 
\frac{2}{M_{\pi}}  \, \sigma_{\pi N}
\,.
\label{eqn:0009c}%
\end{align}
The most recent and precise values for $\sigma_{\pi N}$ are from the Roy-Steiner-equation analyses of
pion-nucleon scattering~\cite{Hoferichter:2015tha,RuizdeElvira:2017stg}. In the calculation with
the inclusion of pionic hydrogen and deuterium data, the reported value is $\sigma_{\pi N}
= (59.1 \pm 3.5)$~MeV, and in the calculation using only the pion-nucleon scattering data the
value is $\sigma_{\pi N} = (58.1 \pm 5)$~MeV. In this study we use the value of
Ref.~\cite{Hoferichter:2015tha} and the uncertainty of Ref.~\cite{RuizdeElvira:2017stg}, which gives 
\begin{align}
x_{1}  =  0.84(7)
\,.
\label{eqn:0009d}%
\end{align}
The parameter $x_{2}$ in Eq.~(\ref{eqn:0009a}) represents the dependence of the strength
of the OPE potential and is given as, 
\begin{align}
x_{2}  = \frac{1}{2F_{\pi}} 
\left. \frac{\partial g_{A}}
{\partial M_{\pi}} \right|_{  M_{\pi}^{\rm ph}}
- \frac{g_{A}}{2F^{2}_{\pi}} 
\left. \frac{\partial F_{\pi}}
{\partial M_{\pi}} \right|_{  M_{\pi}^{\rm ph}}
\,.
\label{eqn:00010a}%
\end{align}
For the dependence of $F_{\pi}$ on $M_{\pi}$ we use the results reported in Ref.~\cite{Berengut:2013nh}
\begin{align}
\left. \frac{\partial F_{\pi}}
{\partial M_{\pi}} \right|_{  M_{\pi}^{\rm ph}}
=
\frac{ F_{\pi}}
{ M_{\pi}}
\frac{K_{F_{\pi}}^{q}}
{K_{M_{\pi}}^{q}}
=0.066(16)
\,.
\label{eqn:00010b}%
\end{align}
The $M_{\pi}$-dependence of the nucleon axial-vector coupling $g_{A}$ is obtained from the analysis of the high-precision lattice QCD calculations~\cite{Chang:2018uxx}. We define 
\begin{align}
\frac{\partial g_{A}}
{\partial M_{\pi}} 
= 
\frac{\partial g_{A}}
{\partial M^{*}} 
\frac{\partial  M^{*}}
{\partial M_{\pi}} \,,
\label{eqn:00010c}%
\end{align}
where
\begin{align}
\frac{\partial  M^{*}}
{\partial M_{\pi}} = 
\frac{\partial }
{\partial M_{\pi}} 
\left(\frac{M_{\pi}}
{4 \pi F_{\pi}}
\right)
= 
\frac{1}
{4 \pi F_{\pi}}
\left(
1 - \frac{M_{\pi}}{F_{\pi}}
\frac{\partial F_{\pi}}
{\partial M_{\pi}}
\right)
=
0.078(2)~{\rm l.u.} \,,
\label{eqn:00010e}
\end{align}
where l.u. stands for lattice units and 
$\left. {\partial g_{A}}/{\partial M^{*}} \right|_{  M_{\pi}^{\rm ph}} = -0.08(24)$.
In Eq.~(\ref{eqn:00010e}) we use the isospin-averaged pion mass $M_{\pi}= 138.03$~MeV.
Putting pieces together, we have  
\begin{align}
\frac{\partial g_{A}}
{\partial M_{\pi}} 
= -0.006(19)~{\rm  l.u.} \,,
\label{eqn:00010g}%
\end{align}
which gives 
\begin{align}
x_{2}  = -0.053(16)~{\rm  l.u.} \,.
\label{eqn:00010h}%
\end{align}

So far we have discussed the quantities $x_{1}$ and $x_{2}$ which control the $M_{\pi}$-dependence of
the pion and nucleon properties as well as their interactions. As has been shown, we obtained a model-independent description of
these quantities utilizing the results from CHPT calculations and the data from high-precision
lattice QCD. Now we turn to the discussion of  the quantities $x_{3}$ and $x_{4}$ which are controlling
the implicit $M_{\pi}$-dependence of the LECs of the short-range NN interactions, $C_0 (M_\pi)$ and
$C_{I}(M_\pi)$. Since  the coupling constants $C_0$ and $C_{I}$ are adjusted to reproduce the NN
scattering phase shifts in the ${}^1S_0$ and ${}^3S_1$ partial waves, it is much more convenient to
express the $x_{3}$ and $x_{4}$ quantities in terms of the inverse singlet ($s$) and triplet ($t$) NN
scattering lengths,
\begin{align}
\bar{A}_s = 
\left. \frac{\partial a_{s}^{-1}}{\partial M_{\pi}} \right|_{  M_{\pi}^{\rm ph}} \,, \quad 
\bar{A}_t = 
\left. \frac{\partial a_{t}^{-1}}{\partial M_{\pi}} \right|_{  M_{\pi}^{\rm ph}} \,.
\label{eqn:0011a}%
\end{align}
To obtain the desired expressions, we adopt the analysis of Ref.~\cite{Epelbaum:2013wla}, which
employs the L\"uscher finite volume formula to relate the spectrum of the NN system in a cubic
periodic box to the NN scattering parameters,
\begin{align}
x_{3}  & = 0.04847 + 0.06713 x_1 - 0.25101 x_2 - 0.37652 \bar{A}_s  - 0.20467 \bar{A}_t \,,\\
x_{4}  & = 0.04990 - 0.00190 x_1 - 0.01253 x_2 - 0.12551 \bar{A}_s  + 0.20467 \bar{A}_t \,.
\label{eqn:0012a}%
\end{align}
We further use the analysis of Ref.~\cite{Lahde:2019yvr}, which determines $\bar{A}_s$ and $\bar{A}_t$
from the most recent available lattice QCD data (see Ref.~\cite{Lahde:2019yvr} for details)
\begin{align}
\bar{A}_s = 
0.54(24) \,, \quad 
\bar{A}_t = 0.33(16) \,.
\label{eqn:0011b}%
\end{align}
Finally, using the results given in Eq.~(\ref{eqn:0011b}) with Eq.~(\ref{eqn:0012a}), we get,
\begin{equation}
x_{3}   = -0.153(96) \,, ~~
x_{4}   =  0.049(46)  \,.
\label{eqn:0012b}%
\end{equation}
In what follows, we will use the values for $\bar{A}_{s,t}$ collected in Eq.~\eqref{eqn:0011b}, noting that
these are still affected by sizeable uncertainties (for a more detailed discussion, see Ref.~\cite{Lahde:2019yvr}).
This can only be sharpened by more precise lattice QCD calculation at lower pion (quark) masses.

\section{Dependence of the nuclear Hamiltonian on the fine-structure constant}
\label{sec:alpha-em-dependence}

First, we must briefly discuss how the electromagnetic interaction is included in our scheme.
This requires a multi-step procedure. In a first step, we consider 8 nucleons (4 protons and
4 neutrons) in a box of $V\simeq (16~{\rm fm})^3$,
from which two $\alpha$ clusters are formed. Here, the EM interaction is included using the standard
power counting, see e.g. \cite{Epelbaum:2005fd}. In this counting, the EM interactions start
to contribute at NLO. To account for the infinitely-ranged Coulomb interaction
between these two clusters with charge $Z=2$ each, we employ a second box of about $V\simeq (100~{\rm fm})^3$,
which is far beyond the range of the strong interactions. Within this box, a spherical wall with a
radius of about $35$~fm is placed subject to Coulomb boundary conditions. This allows for an exact
treatment of the long-range Coulomb forces with the two $\alpha$ particles. For details on this procedure,
we refer to Refs.~\cite{Elhatisari:2015iga,Elhatisari:2016hby}.

Now,  we are in the position to consider the second term on the right-hand side of Eq.~(\ref{eqn:0001}), which is
the $\alpha_{\rm EM}$-dependence of $\alpha$-$\alpha$ scattering. To study the $\alpha_{\rm EM}$-dependence
of $\alpha$-$\alpha$ scattering we compute the shifts $\Delta E_{\alpha\alpha}(\alpha_{\rm EM})$ and
$\Delta E_{\alpha\alpha}(c_{pp})$. The former is the variation of two-alpha cluster energy due to the
long-range Coulomb interaction, and the latter
is the variation of two-alpha cluster energy due
to a derivative-less proton-proton contact operator. This operator arises from the fact that the
Coulomb interaction on the lattice becomes singular when two protons are on the same lattice site
which requires a special treatment. Thus, a regularized version of the Coulomb interaction on the
lattice is employed, and the coefficient of the proton-proton contact operator, $c_{pp}$, is
determined from the proton-proton phase shifts on the lattice. The energy shift becomes,
\begin{align}
Q_{\rm EM} (E_{\alpha\alpha}) =  \Delta E_{\alpha\alpha}(\alpha_{\rm EM})
+x_{pp} \, \Delta E_{\alpha\alpha}(c_{pp}) \,,
\label{eqn:0020a}%
\end{align}
where $x_{pp}$ is the relative strength of the proton-proton contact term caused by the regularization of
the Coulomb force. The coefficient  $x_{pp}$  is computed using the data for $^4$He~\cite{Epelbaum:2013wla},
\begin{align}
x_{pp} = 0.39(5)\,.
\label{eqn:0020b}%
\end{align}
Finally, the partial derivative in Eq.~(\ref{eqn:0001}) can be written as
\begin{align}
\left. \frac{\partial E_{\alpha\alpha}}
{\partial \alpha_{\rm EM}}  \right|_{\alpha_{\rm EM}^{\rm ph}} 
\simeq
\frac{\partial Q_{\rm EM} (E_{\alpha\alpha})}
{\partial \alpha_{\rm EM}^{\rm ph}}\,.
\label{eqn:0020c}%
\end{align}

\section{Theta-dependence of alpha-alpha scattering}
\label{sec:theta}

We also strive to assess the $\theta$-dependence of $\alpha$-$\alpha$ scattering. To that end, one might be
tempted to again employ a linear variation $\propto\delta\theta$ around the physical
value of $\theta^{\rm ph}$, similar to what we do in the case of the $M_\pi$- and the
$\alpha_{\rm EM}$-dependence of $E_{\alpha\alpha}$, see Eq.~(\ref{eqn:0001}). However, it is well known that
``small'' variations of $\theta$ do not lead to drastic changes of nuclear physics
\cite{Ubaldi:2008nf,Lee:2020tmi} and after all it is interesting in its own right to assess
what is happening when $\theta$ approaches a value of, say, $\mathcal{O}(1)$. In this regime, a
simple linear variation clearly would not be applicable any longer.

There is, however, a way to circumvent such a direct calculation of the $\theta$-dependence, which
is based on the observation that in a first approximation any source of $\theta$-dependence of
$E_{\alpha\alpha}$ can be traced back to the $\theta$-dependence of $M_\pi$, which in the isospin
limit is given by \cite{Brower:2003yx}\footnote{The physics at $\theta=\pi$ is a bit more involved,
see, e.g., \cite{Smilga:1998dh,Vonk:2019kwv}.}
\begin{equation}\label{eq:mpi}
M_\pi^2(\theta) = 2 B_0 \hat{m} \, \cos\frac{\theta}{2}\, , ~|\theta| < \pi \, .
\end{equation}
Assuming this approximation is valid, the present calculation of the $M_\pi$-dependence of $E_{\alpha\alpha}$
within a range of $|\delta M_\pi| \lesssim 0.1 M_\pi^{\rm ph}$ can directly be translated into
an assessment of the $\theta$-dependence in a corresponding range of $|\delta \theta| \lesssim 1$.

It is not obvious that 
this approximation is legitimate, as $M_\pi$ and $\theta$ in CHPT are in principle
independent parameters, but it can be justified as follows: Removing the QCD $\theta$-term by a suitable
choice of an axial U(1) transformation adds a complex phase 
\begin{equation}
\mathcal{M} \to e^{\im \frac{\theta}{2}}
\mathcal{M} =: \mathcal{M}_\theta,
\end{equation}
to the quark mass matrix.
This $\theta$-dependent matrix enters chiral perturbation theory via the matrix $\chi_\theta = 2
B_0 \mathcal{M}_\theta$, which in the isospin symmetric case is simply given by
\begin{equation}\label{eq:massmatrix}
\mathcal{\chi}_\theta = \left(M_\pi^2(\theta)+\im\,2 B_0 \hat{m} \sin\frac{\theta}{2}\right)\mathbbm{1}\, .
\end{equation}
Hence, 
inserting this expression into a given chiral Lagrangian of any order will produce terms that
are either proportional to (some power of) $M_\pi(\theta)$, or proportional to (some power of)
$\sin\theta/2$ (or both). While the latter are naturally absent in chiral perturbation theory
at $\theta=0$, the former simply leads to the known $M_\pi$-dependence of quantities such as
$m_N$, $\tilde{g}_{\pi N}$, or couplings of nucleons to two or more pions.

As it turns out, at NLO, which is the maximal order we are considering here, the only term
$\propto \sin\theta/2$ that might alter any of the involved quantities, in particular
$m_N$ or $\tilde{g}_{\pi N}$, comes from the NLO pion-nucleon Lagrangian \cite{Bernard:1996gq}
\begin{equation}
\mathcal{L}^{(2)}_{\pi N} = c_5 \bar{N} \left(\chi_+ -\frac{1}{2}\Tr \chi_+  \right) N + \dots\, ,
\end{equation}
where $c_5$ is another LEC, the ellipses represent other NLO terms that are of no interest here, and 
\begin{equation}
\chi_+ = u^\dagger \chi u^\dagger + u \chi^\dagger u\, ,
\end{equation}
with $u$ carrying the pion fields. This term adds a contribution to the pion-nucleon coupling that
is explicitly $\theta$-dependent, but it can be shown that its actual numerical impact is so
small ($\lesssim$ 1–2\%)  \cite{Ubaldi:2008nf,TVmaster} that it can safely be neglected.
The smallness of these effects can directly be traced back to the suppression of the
LEC $c_5 = (-0.09\pm 0.01)\,$GeV$^{-1}$ as it parameterizes the leading isospin-breaking
effects in the pion-nucleon sector \cite{Bernard:1996gq}.
This means that as long as we stick to a calculation that is of NLO at most,
any non-negligible $\theta$-dependence indeed only appears implicitly in form of $M_\pi(\theta)$ as a
consequence of the first term of Eq.~\eqref{eq:massmatrix}.

Thus, our approach here is to not perform a separate calculation for assessing the
$\theta$-dependence of $\alpha$-$\alpha$ scattering, but to simply use the results of the
$M_\pi$-dependence analysis and map them onto the $\theta$-dependence using Eq.~\eqref{eq:mpi}.

\section{Adiabatic Projection Method and Auxiliary Field Quantum Monte Carlo Simulations}
\label{sec:tools}

The adiabatic projection method is a general framework to construct a low-energy effective theory
for clusters. The adiabatic projection in Euclidean time gives a systematically improvable
description of the low-lying scattering cluster states and in the limit of large Euclidean
projection time the description becomes exact. The details of the method can be found in
Refs.~\cite{Pine:2013zja,Elhatisari:2016hby}.  The method starts with defining Slater-determinant of
two-alpha initial cluster states $\ket{\vec{R}}$ parameterized by the relative spatial
separation between the clusters on a periodic cubic lattice with a box size $L$,
\begin{align}
\ket{\vec{R}} = \sum_{\vec{r}} \ket{\vec{r}+\vec{R}}_{1}\otimes \ket{\vec{r}}_{2}\,.
\label{eqn:0017}%
\end{align}
To perform the calculations efficiently, we project the initial states onto spherical harmonics
with angular momentum quantum numbers $\ell$ and $\ell_z$. To that end, we bin the cubic
lattice points $\braket{n_{x},n_{y},n_{z}}$ with the same distance $|\vec{R}| = \sqrt{n_{x}^2 + n_{y}^2
+ n_{z}^2}$ by weighting with spherical harmonics $Y_{\ell,\ell_z}(\hat{R})$, 
\begin{align}
\ket{R}^{\ell,\ell_z} = \sum_{\vec{r}} Y_{\ell,\ell_z}(\hat{R}^{\prime}) \delta_{\vec{R},|\vec{R}^{\prime}|}\ket{\vec{R}}\,.
\label{eqn:0017a}%
\end{align}
Here, $\ell$ and $\ell_z$ are not exactly good quantum numbers, see the discussion
in Ref.~\cite{Lu:2015riz}.
Since the initial cluster states are not necessarily orthonormal, we define the orthonormal initial
cluster states 
\begin{align}
\ket{\mathcal{R}}^{\ell,\ell_z} = 
 \sum_{R^{\prime}} \ket{R^{\prime}}^{\ell,\ell_z}
  [N_{0}^{-1/2}]^{\ell,\ell_z}_{R^{\prime},R}
 \,,
\label{eqn:0017b}%
\end{align}
where $[N_{0}^{-1}]^{\ell,\ell_z}_{R^{\prime},R}$ is the norm matrix defined as
\begin{align}
[N_{0}^{-1}]^{\ell,\ell_z}_{R^{\prime},R} = {}^{^{\ell,\ell_z}}\!\braket{R^{\prime}  | R}^{^{\ell,\ell_z}}\,.
\label{eqn:0017c}%
\end{align}
In the next step, the initial cluster states are evolved in Euclidean time by means of multiplying
by powers of the leading order (LO) transfer matrix to form dressed cluster states,
\begin{align}
\ket{R}^{\ell,\ell_z}_{n_{t}} = M_{\rm LO}^{n_{t}} \ket{\mathcal{R}}^{\ell,\ell_z}\,.
\label{eqn:0021a}%
\end{align}
This procedure, by design, incorporates all the induced  deformations and polarizations of the
alpha clusters due to the microscopic interaction and it gives the true low-lying cluster states
of the transfer matrix $M_{\rm LO}$. In general the dressed cluster states are not orthonormal,
thus for further calculations we use the following form of the dressed cluster states, 
\begin{align}
\ket{\mathcal{R}}^{\ell,\ell_z}_{n_{t}} = 
\sum_{R^{\prime}}
 \ket{R^{\prime}}^{\ell,\ell_z}_{n_{t}}
 \,\,
  [N_{L_t}^{-1/2}]^{\ell,\ell_z}_{R^{\prime},R}
 \, ,
\label{eqn:0021b}%
\end{align}
where $[N_{L_t}^{-1/2}]^{\ell,\ell_z}_{R,R^{\prime}}$ is the norm matrix at Euclidean time $L_{t} = 2 \times n_{t}$.
Finally, we define the radial adiabatic transfer matrix at LO as,
\begin{align}
[M^{a}_{{\rm LO}, L_{t}}]^{\ell,\ell_z}_{R,R^{\prime}}  =
\,\,
{}^{\ell,\ell_z}_{n_{t}}\!\bra{\mathcal{R}}
M_{{\rm LO}}
\ket{\mathcal{R}^{\prime}}^{\ell,\ell_z}_{n_{t}}
 \,.
\label{eqn:0021e}%
\end{align}

In our calculation the higher-order interactions are treated using first-order perturbation theory,
thus we include the perturbative contributions from NLO, next-to-next-to-leading order (NNLO),
isospin-breaking (IB), and Coulomb
interactions (EM) to the leading-order radial adiabatic transfer matrix  order-by-order in perturbation
theory. Therefore, we define the radial adiabatic transfer matrix at a given higher order in
a closed form as
\begin{align}
[M^{a}_{{\rm HO}, L_{t}}]^{\ell,\ell_z}_{R,R^{\prime}}  =
\,\:
{}^{\ell,\ell_z}_{n_{t}}\!\bra{\mathcal{R}}
M_{{\rm LO}}
\ket{\mathcal{R}^{\prime}}^{\ell,\ell_z}_{n_{t}}
-
\alpha_{t} \,\:
{}^{\ell,\ell_z}_{n_{t}}\!\bra{\mathcal{R}}
: V_{\rm HO} \, M_{{\rm LO}} :
\ket{\mathcal{R}^{\prime}}^{\ell,\ell_z}_{n_{t}}
\,,
\label{eqn:0025}%
\end{align}
where $\alpha_{t} = a_{t}/a$ is the ratio of the temporal and the spatial lattice spacings, and
$V_{\rm HO}$ is the higher-potential at the order of interest.  The colons $:...:$ denote normal ordering, which means that we reorder the creation and annihilation operators inside the colons and we move the creation operators to the left of the all annihilation operators with the appropriate number of anti-commutation minus signs.

So far we have discussed the adiabatic projection method for the chiral EFT Hamiltonian. Now we
turn to the main interest of this paper, which is to construct the two-cluster matrix elements of
the partial derivatives given in \eref{eqn:0001}. Due to the fact that we study the effects of small
variations in the fundamental constants of nature  on $\alpha$-$\alpha$ scattering, the partial derivatives
in \eref{eqn:0001} are treated in a similar manner as the higher-order corrections,
\begin{align}
[M^{a,y}_{{\rm HO}, L_{t}}]^{\ell,\ell_z}_{R,R^{\prime}}
 =
 &
\,\:
{}^{\ell,\ell_z}_{n_{t}}\!\bra{\mathcal{R}}
M_{{\rm LO}}
\ket{\mathcal{R}^{\prime}}^{\ell,\ell_z}_{n_{t}}
\nonumber\\
&
-
\alpha_{t} \,\:
{}^{\ell,\ell_z}_{n_{t}}\!\bra{\mathcal{R}}
: V_{\rm HO} \, M_{{\rm LO}} :
\ket{\mathcal{R}^{\prime}}^{\ell,\ell_z}_{n_{t}}
\nonumber\\
&
-
\alpha_{t} \,\:
{}^{\ell,\ell_z}_{n_{t}}\!\bra{\mathcal{R}}
: 
\left. \frac{\partial E_{\alpha\alpha}}
{\partial y}          \right|_{y^{\rm ph}}
\, M_{{\rm LO}} :
\ket{\mathcal{R}^{\prime}}^{\ell,\ell_z}_{n_{t}} \, \,  \delta y
\,,
\label{eqn:0027}%
\end{align}
we use the superscript $y$ for the observables $M_{\pi}$, $\alpha_{\rm EM}$ and, in principle, $\theta$.
However, as discussed in Sec.~\ref{sec:theta}, we will not perform explicit differentiations with
respect to $\theta$.

The two-cluster matrix elements of the LO transfer matrix, the higher order corrections, and the
partial derivatives are computed by means of the auxiliary field quantum Monte Carlo (AFQMC) method.
The non-perturbative quantum Monte Carlo simulations are performed using the neutral pion mass $M_{\pi}$
and the isospin symmetry breaking effects are incorporated perturbatively. The calculation of the radial
adiabatic transfer matrices in Eqs.~(\ref{eqn:0021e}), (\ref{eqn:0025}) and (\ref{eqn:0027}) is divided
into two separate parts. In the first part of the calculation, we perform the AFQMC simulation for
the system of $A = 8$ nucleons (4 protons and 4 neutrons) to construct the radial adiabatic transfer
matrices for two interacting $\alpha$ clusters. Due to the computational cost associated with such
simulations, this is done on a periodic cubic lattice of length $L$ which is not too large to
prevent us from computing the matrices accurately but is not too small so that the length $L/2$ is much
larger than the range of the interaction, $R\sim 1/M_\pi \simeq 1.4\,$fm. In the second part of the
calculation, the AFQMC simulations
are performed for the system with $A = 4$ nucleons, and these simulations are done on a periodic cubic
lattice of larger length due to the less computational demand. This single $\alpha$ cluster adiabatic
matrix is used to construct the radial adiabatic transfer matrices for non-interacting two $\alpha$
clusters. Finally, we connect the radial adiabatic transfer matrices of interacting $\alpha$
clusters with the radial adiabatic transfer matrices of non-interacting $\alpha$ clusters in the
asymptotic region to extend the radial transfer matrix of interacting $\alpha$ clusters to a larger
volume. The aforementioned two-part approach was studied extensively for nucleon-deuteron systems
in Ref.~\cite{Elhatisari:2016hby}, and it was found that the systematic errors due to extension of
the radial transfer matrix are negligible.

The first \textit{ab initio} calculation of $\alpha$-$\alpha$ scattering was performed in
Ref.~\cite{Elhatisari:2015iga} using the same chiral Hamiltonian as adopted in this paper.
However, in this paper we employ developments in the adiabatic projection method from
Refs.~\cite{Elhatisari:2016hby,Elhatisari:2016owd,Elhatisari:2019fvk}. As discussed above, the first
step of the adiabatic projection method is to define the initial cluster states, and on a periodic
cubic lattice of length $L$ the total number of initial cluster states parameterized by the
relative spatial separation is $N_{\vec R} = 3L^2/4$. In Ref.~\cite{Pine:2013zja} it was shown that
it is not required to use every possible cluster state when we are interested in only a few low-lying
energies of the system of interest. Therefore, for simulating computationally demanding systems it
is advantageous to construct a radial adiabatic transfer matrix defined in the subspace that
is spanned by $N_{\vec R} < 3L^2/4$ cluster separation states. Following these findings,
in Ref.~\cite{Elhatisari:2015iga} the radial adiabatic transfer matrix for non-interacting two-alpha
clusters was constructed in a smaller subspace of the two-cluster state space. In this paper,
taking advantage of powerful computational resources we perform our simulations using every possible
cluster state and construct the  radial adiabatic transfer matrices in full space of the two-cluster
state space.

\section{Extracting Scattering Phase Shifts from the Adiabatic Matrices}
\label{sec:APM-SPS}

What was discussed in the previous section was  the first part of the adiabatic projection
method, which is constructing the adiabatic transfer matrix for the two clusters. The second part of the
method is to extract the scattering or reaction parameters for the two clusters. In the previous
section, by projecting the initial cluster states onto spherical harmonics with angular momentum quantum
numbers $\ell$ and $\ell_z$, we constructed the adiabatic transfer matrix in radial coordinates, which
provides a significant improvement in the computational scaling~\cite{Elhatisari:2016hby}. Since our
adiabatic transfer matrices are defined in  radial coordinates, the best approach to be used
to calculate the scattering parameters is the so-called spherical wall
method~\cite{Carlson:1984zz,Borasoy:2007vy,Lu:2015riz}.

In the spherical wall method we employ a hard boundary wall condition at $r = R_{\rm wall}$, which is the
relative separation distance between two clusters in the asymptotic region. In general, the spherical wall
method is used to remove the periodic boundary effects inherited from the cubic lattice and the
artifacts due to the periodic boundary condition. However, in our calculations these effects are already
eliminated since we construct the adiabatic transfer matrices in radial coordinates as explained in
Sec.~\ref{sec:tools}. After imposing the spherical hard wall to the radial adiabatic transfer matrices,
we solve the Schr\"odinger equation of the system and obtain the spherical scattering wave functions as
well as the spectrum. In principle, due to the imposed spherical hard wall one expects that the
spherical wave functions die out at $R_{\rm wall}$, however, as a result of non-zero spatial lattice spacing
the spherical wave functions vanish at $R_{\rm wall}^{\prime} = R_{\rm wall}+\varepsilon_{R}$, where $\varepsilon_{R}$
is the correction on the precise radius of the spherical wall and is defined as $|\varepsilon_{R}|<a/2$.

The total wave function of a two-cluster system is decomposed into the radial part $R_{\ell}^{(p)}(r)$
and the spherical harmonics $Y_{\ell,\ell_z}^{}(\hat{r})$,
\begin{align}
\Psi(\vec{r}) =  R_{\ell}^{(p)}(r)\,Y_{\ell,\ell_z}^{}(\hat{r})\,,
\label{eqn:0030}%
\end{align}
where $r$ is the relative spatial separation of the clusters and $p$ is the relative momentum. The radial
wave function in the asymptotic region is given by
\begin{align}
R_{\ell}^{(p)}(r)
= 
N_{\ell}(p) 
\left[
\cos\delta_{\ell}(p) \, F_{\ell}(p \, r) 
+
\sin\delta_{\ell}(p) \, G_{\ell}(p \, r)
\right]
\,,
\label{eqn:0033}%
\end{align}
where $N_{\ell}(p)$ is an overall normalization coefficient, and $F_{\ell}~(G_{\ell})$ is the regular (irregular)
Coulomb wave function.

The relative momentum $p$ is calculated from the spectrum of the radial adiabatic transfer matrices
and the dispersion relation of the two-cluster system given by,
\begin{align}
E = c_{0} \, \frac{p^2}{2\mu} + c_{1} \, p^4 + c_{2} \, p^6 + \ldots
\,,
\label{eqn:0035}%
\end{align}
where $\mu =m_{\alpha}/2$ is the reduced mass of the two-cluster system, $m_\alpha$ the mass of the
$\alpha$-particle, and the coefficients $c_{i}$ are determined by
fitting Eq.~(\ref{eqn:0035}) to the lattice dispersion relation. We determine the correction $\varepsilon_{R}$
from the roots of the regular Coulomb wave function with the relative momentum of the non-interacting
two-cluster system, $p_{0}$, around $R_{\rm wall}$. Finally, we use the corrected radius of the spherical
hard wall, $R_{\rm wall}^{\prime}$, and the relative momentum of the interacting two-cluster system, $p$,
and solve Eq.~(\ref{eqn:0033}) for the scattering phase shifts, 
\begin{align}
\delta_{\ell}(p) = \tan^{-1}
\left[-
\frac{F_{\ell}(p \, R_{\rm wall}^{\prime})}
{G_{\ell}(p \, R_{\rm wall}^{\prime})}\,  
\right]
\,.
\label{eqn:0033a}%
\end{align}
We extract the scattering phase shifts from the radial adiabatic transfer matrices with $L_t$ time steps and perform Euclidean time extrapolating to the limit $L_{t} \to \infty$. Details of the extrapolation fit and
all associated error estimates are discussed in Appendix~\ref{app:Ete}.


\section{Results}
\label{sec:results}

\subsection{Our universe}
\label{sec:novar}

\begin{figure}[htb]
\includegraphics[width=0.48\textwidth]{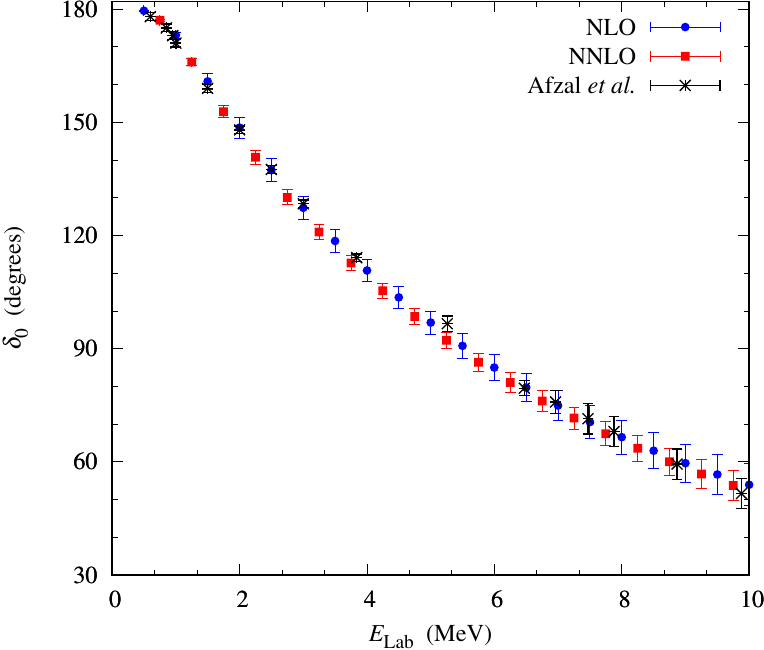}~~
\includegraphics[width=0.48\textwidth]{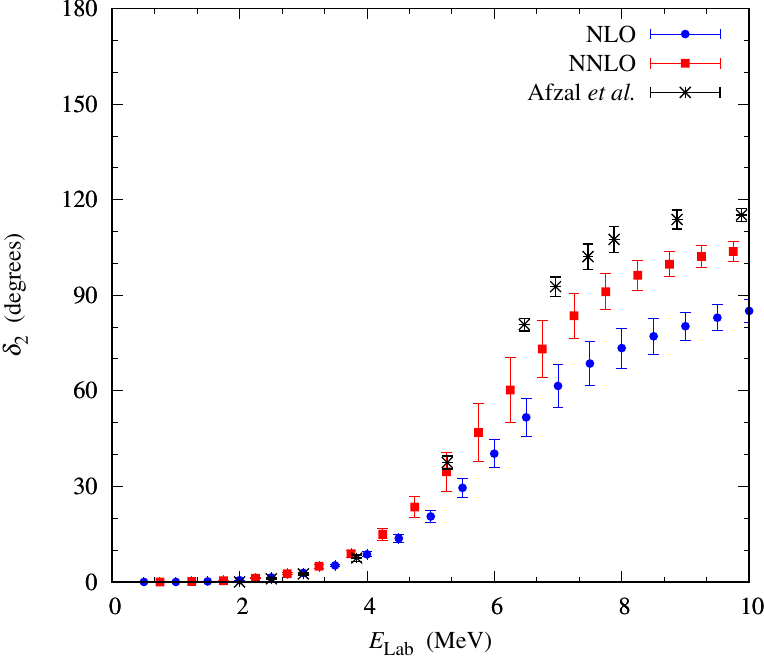}
\caption{Left panel: S-wave $\alpha$-$\alpha$ scattering phase shift $\delta_0$
versus the energy in the laboratory system, $E_{\rm lab}$.
Right panel: D-wave $\alpha$-$\alpha$ scattering phase
shift $\delta_2$ versus the energy in the laboratory system, $E_{\rm lab}$. 
The blue circles and red squares represent our predictions at NLO and NNLO, respectively, while the data
are given by the black crosses.}
\label{fig:phases_novar}
\end{figure}

Here, we discuss the results for the S- and D-wave phase shifts and the effective range
parameters in the S-wave as well as the resonance parameters in the D-wave
for the physical values of $M_\pi$ and $\alpha_{\rm EM}$ and $\theta = 0$. In
Fig.~\ref{fig:phases_novar}, we show the S-wave phase shift $\delta_0$ (left panel) and the D-wave
phase shift $\delta_2$ (right panel) at NLO and NNLO in comparison to  the data \cite{Afzal:1969hal}.
Note that we do not show the LO result here, as the electromagnetic interaction is not yet included
and therefore the predicted curve is far off the data (as discussed in more detail in
Ref.~\cite{Elhatisari:2015iga}). We find a marked improvement, both for the S-wave and the D-wave,
as compared to the pioneering work in Ref.~\cite{Elhatisari:2015iga}, which is due to the
improvements in the APM discussed in the earlier sections. We note that these are
parameter-free predictions. Furthermore, the uncertainties are mostly stemming from the
large Euclidean time extrapolation and these decrease when going from NLO to NNLO, as
expected in a well-behaved expansion. 
Up to $E_{\rm Lab}\simeq 3.5\,$MeV, our description of the S-wave phase shift is as good
as the one obtained using halo EFT in Ref.~\cite{Higa:2008dn}. We note that the
uncertainties have somewhat increased as compared to Ref.~\cite{Elhatisari:2015iga}
because, as discussed in the previous section, we use a much larger subspace of the two-cluster
state space, which reduces the number of configurations used for the matrix entries,
resulting in a larger statistical uncertainty. This could eventually be overcome by utilizing
much more HPC resources.

Next, we discuss the S-wave ERE parameters $a_0$, $r_0$ and $P_0$ (see Appendix~\ref{app:modERE}
for definitions), collected in Table~\ref{table:EREplus}. The fit range to determine these
is from $E_{\rm Lab}=1.0$ to $7.7$~MeV. We see that these parameters are consistent with the empirical determinations,
but they are also afflicted with sizeable uncertainties. Note that there is sensitivity to the
fit range as well as to the position of the $0^+$ resonance, the $^8$Be ground state,
as discussed in Ref.~\cite{Higa:2008dn}. In our calculation, $^8$Be is very weakly bound. This appears to
be in contradiction to the scattering lengths given in Tab.~\ref{table:EREplus}, but these values
are very sensitive to the fitting range employed to extract them, see also Ref.~\cite{Higa:2008dn}.

\begin{table}[htb]
  \caption{S-wave: The ERE parameters $a_0, r_0$ and $P_0$ at NLO and NNLO.
    D-wave: The resonance parameters $E_R$ and $\Gamma_R$ at NLO and NNLO.
    The empirical values from Ref.~\cite{Rasche:1967urp} are also given.
  }
\label{table:EREplus}
\centering
\begin{tabular}{|l|c|c|c||c|c|}
\hline  
  & \multicolumn{3}{c||}{S-wave} &  \multicolumn{2}{c|}{D-wave}  \\
  \hline
& $a_{0}$~[$10^3$~fm]   & $r_{0}$~[fm]&   $P_{0}$~[fm$^3$] & $E_R$~[MeV] & $\Gamma_R$~[MeV]\\
  \hline
NLO        &  $-$1.80(93)  & 1.045(15)   &$-$2.297(156) &  3.05(4)  & 2.68(23)\\
  \hline
NNLO       &  $-$1.55(63)  & 1.061(14)   &$-$2.277(158)&  2.93(5)  & 2.00(16)\\
  \hline
empirical  &  $-$1.65(17)  & 1.084(11)   &$-$1.76(22)&  2.92(18)  & 1.35(50)\\
  \hline
\end{tabular}
\end{table}

The D-wave phase shift shows a clear resonance-behaviour. Due to the large width of the
resonance, the extraction of the resonance parameters (energy and width) is affected with
some model-dependence. As in our earlier work, we fix the resonance energy $E_R$ by the
maximum of $d\delta/dE$ and its width $\Gamma_R$ from the value of $2(d\delta/dE)^{-1}$ at
$E_R$, see e.g. Ref.~\cite{Hupin:2014kha}. The resonance parameters at NLO and NNLO are
also given in Tab.~\ref{table:EREplus}. We find that the resonance parameters at NNLO
are much closer to the empirical ones as compared to our earlier work.

\subsection{The Multiverse}
\label{sec:var}

\subsubsection{Variations of the bound state energies}
\label{sec:energ}

\begin{figure}[tb]
\includegraphics[width=0.9\textwidth]{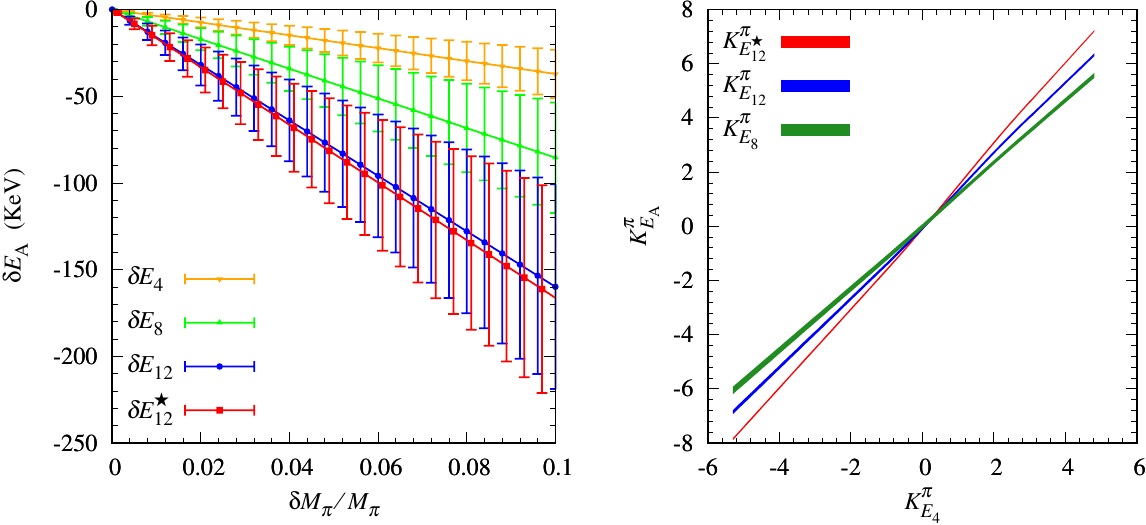}
\caption{Left panel: Variation of the ground state energy of the nuclei $^4$He,  $^4$Be, $^{12}$C
and the Hoyle state $^{12}$C$^\star$, respectively, under variation of the pion mass (in percent).
Right panel: Sensitivity of the ground state energy of the nuclei $^8$Be, $^{12}$C
and the Hoyle state $^{12}$C$^\star$, respectively, to changes in $M_\pi$ as a function of $K_{E_4}^\pi$
under independent variations of $\bar{A}_s$ and $\bar{A}_t$ over the range $\{-1, \ldots,+1\}$.
}
\label{fig:bs}
\end{figure}

Before considering the effect of the variations of the fundamental parameters on the $\alpha$-$\alpha$
scattering phase shifts, we discuss briefly the variation of the various bound state energies
relevant to the $3\alpha$ process. This provides some additional information to Ref.~\cite{Lahde:2019yvr}
that was not explicitly displayed there. Consider first pion mass variations, keeping $\alpha_{\rm EM}$
and $\theta$ at their physical values. In the left panel of Fig.~\ref{fig:bs}, we display the variation
of the energies of $^4$He, $^4$Be, $^{12}$C and the Hoyle state $^{12}$C$^\star$ as a function of the
varying pion mass for positive changes in $M_\pi$.
These energies are denoted as $E_4$, $E_8$, $E_{12}$ and $E_{12}^\star$, in order, see
the explicit expressions in App.~\ref{app:bsmpi}. For negative
energy changes in $M_\pi$ these curves only differ in the sign, that is the contribution
is repulsive rather than attractive as for positive shifts in the pion mass.
These different energies are obviously correlated,
as shown more clearly in the right panel of  Fig.~\ref{fig:bs}, where the various $K$-factors 
for the pertinent eight and twelve particle systems are displayed as a function of the corresponding
$^4$He $K$-factor, $K_{E_4}^\pi$, for independent variations of $\bar{A}_s$ and $\bar{A}_t$ over the
range $\{-1, \ldots,+1\}$ are shown. Of course, the actual range of these parameters as given
in Eq.~\eqref{eqn:0011b} is smaller, but these parameters might change when better results from
lattice QCD will become available. Note that such correlations related to the production of
carbon have indeed been speculated upon earlier~\cite{Livio,WeinbergFacing}.

\begin{figure}[tb]
\centering\includegraphics[width=0.50\textwidth]{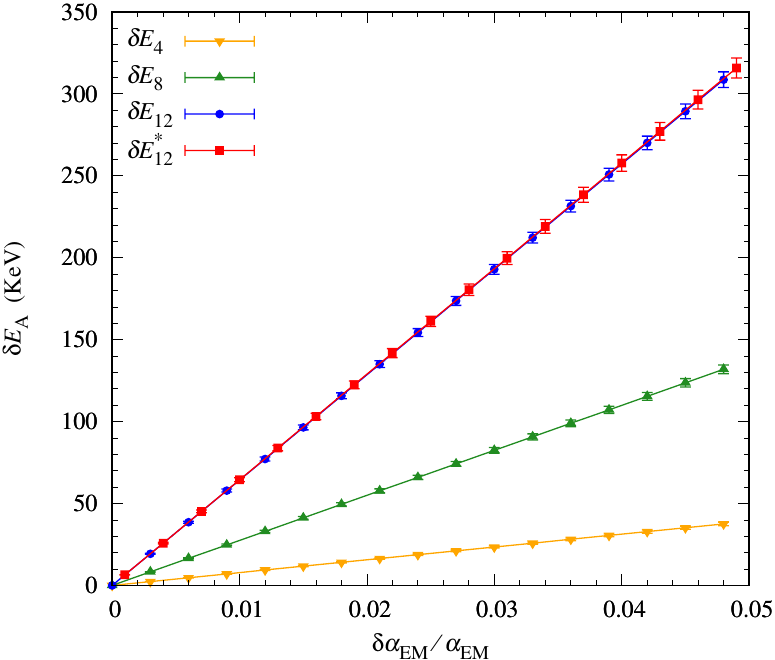}
\caption{Variation of the ground state energy of the nuclei $^4$He,  $^4$Be, $^{12}$C
and the Hoyle state $^{12}$C$^\star$, respectively, under variation of the
fine-structure constant (in percent).}
\label{fig:bs_aEM}
\end{figure}

Next, we consider variations of the fine-structure constant for physical pion masses and vanishing
$\theta$ angle. The variations of the energies  $E_4$, $E_8$, $E_{12}$ and $E_{12}^\star$ with varying
$\alpha_{\rm EM}$ are displayed in Fig.~\ref{fig:bs_aEM} (for positive shifts in $\alpha_{\rm EM}$).
Naively, one would expect the slopes of the different nuclei to scale as $Z^2$, that is in the
ratio $1:4:9$ for $^4$He, $^8$Be and $^{12}$C, in order. The observed difference from this
scaling is coming from the proton-proton derivative-less contact interaction. In fact, removing
the contribution from this term, one finds for $\delta \alpha_{\rm EM}/\alpha_{\rm EM} = 5\%$ the following
energy shifts: $\delta E_4 = 30.65(10)$~keV, $\delta E_8 = 117.5(10)$~keV and  $\delta E_{12} = 283.5(10)\,$keV,
perfectly consistent with the $Z^2$ scaling. We note here that the results for negative shifts
in $\alpha_{\rm EM}$ are of opposite sign, that is pertinent  energy shifts $\delta E_A$ are negative.

\subsubsection{Pion mass variations of alpha-alpha scattering}
\label{sec:mpi}

\begin{figure}[htb]
\includegraphics[width=0.48\textwidth]{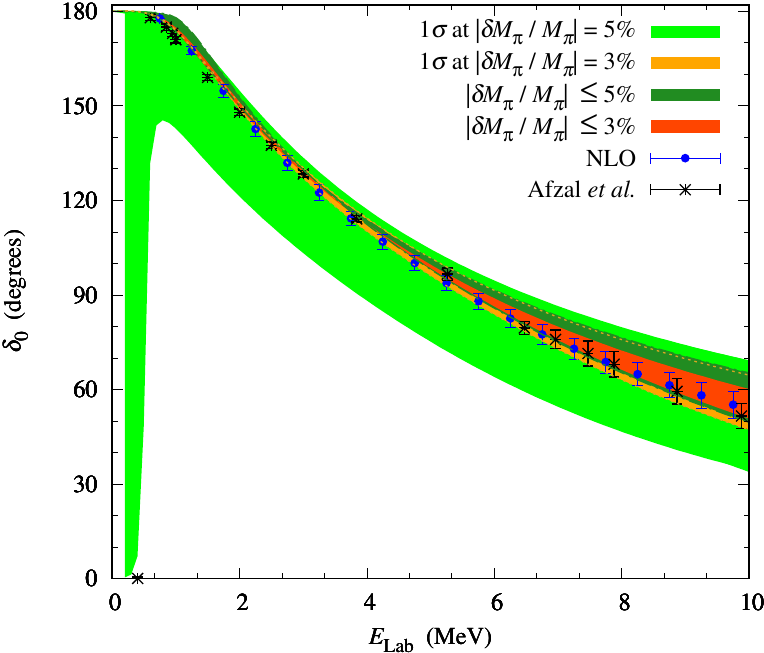}~~
\includegraphics[width=0.48\textwidth]{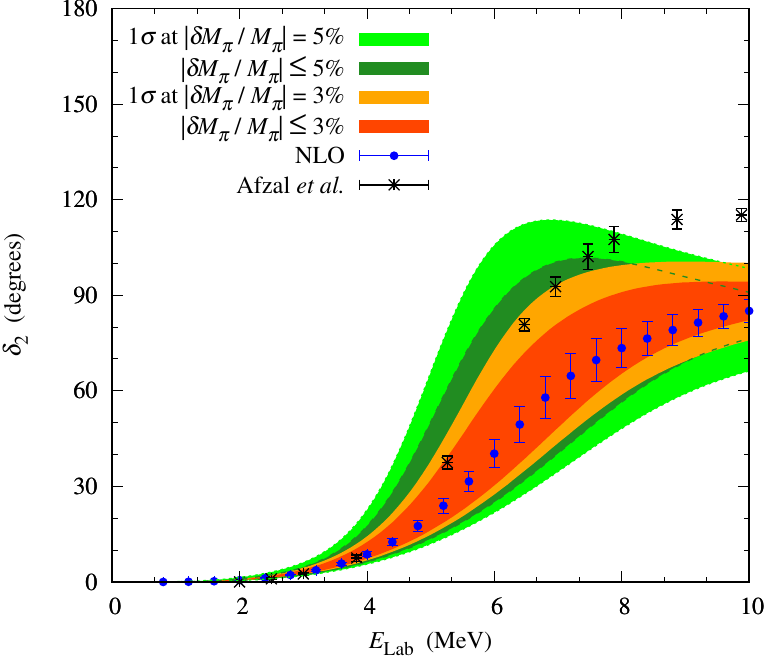}
\caption{Pion mass dependence of the $\alpha$-$\alpha$ phase shifts at NLO.
Left panel: S-wave  phase shift $\delta_0$
versus the energy in the laboratory system, $E_{\rm lab}$.
Right panel: D-wave  phase
shift $\delta_0$ versus the energy in the laboratory system, $E_{\rm lab}$.
The black crosses refer to the exprimental data, the blue circles  are the NLO results
in the limit $L_{t}\to \infty$  at $\delta M_{\pi}=0$. The red band corresponds the S-wave
phase shifts with a variation in $M_{\pi}$ within $|\delta M_{\pi}/M_{\pi}|\leq 3\%$, and the
golden band refers to the errors for $|\delta M_{\pi}/M_{\pi}|=3\%$. The dark green
band corresponds to a variation in $M_{\pi}$ within $|\delta M_{\pi}/M_{\pi}| \leq 5\%$,
and the light green band refers to the  errors for $|\delta M_{\pi}/M_{\pi}| =5\%$.
In the case of variation in $M_{\pi}$ by $-5\%$ in the S-wave, due to difficulty in performing Euclidean time
extrapolation at low-energies we estimate the error band from the spread in phase shifts versus
the number of time steps.}
\label{fig:phases_mpi_NLO}
\end{figure}

\begin{figure}[htb]
\includegraphics[width=0.48\textwidth]{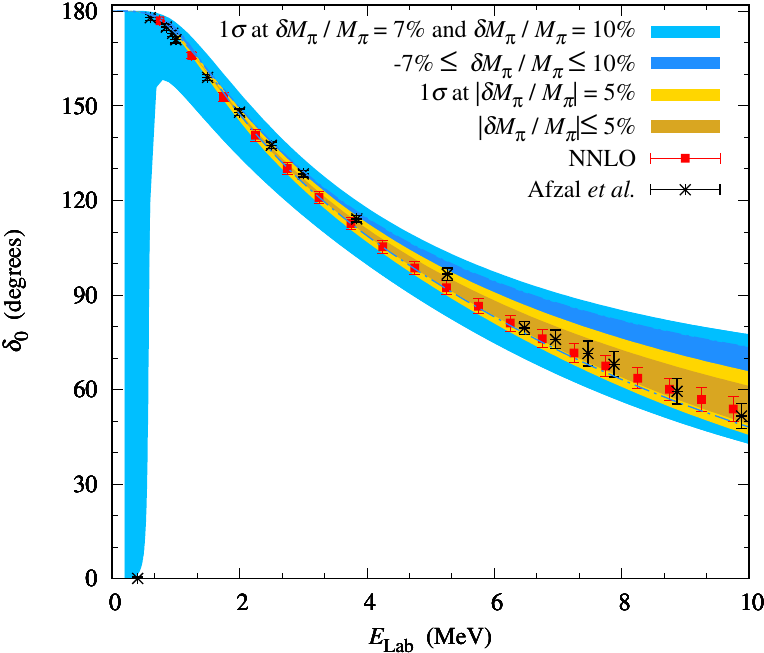}~~
\includegraphics[width=0.48\textwidth]{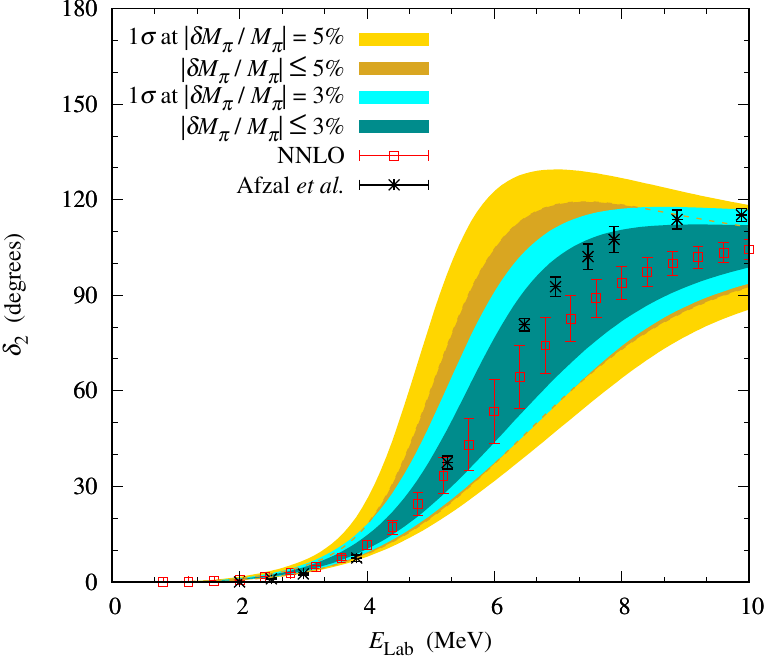}
\caption{Pion mass dependence of the $\alpha$-$\alpha$ phase shifts at NNLO.
Left panel: S-wave  phase shift $\delta_0$
versus the energy in the laboratory system, $E_{\rm lab}$.
Right panel: D-wave  phase
shift $\delta_0$ versus the energy in the laboratory system, $E_{\rm lab}$.
The black crosses refer to the experimental data, the red squares are the NNLO results
in the limit $L_{t}\to \infty$  at $\delta M_{\pi}=0$. The dark gold band corresponds to the S-wave
phase shifts with a variation in $M_{\pi}$ within $|\delta M_{\pi}/M_{\pi}|\leq 5\%$, and the
light golden band refers to the errors for $|\delta M_{\pi}/M_{\pi}|=5\%$. The dark blue
band corresponds to a variation in $M_{\pi}$ within $-7\% \leq \delta M_{\pi}/M_{\pi} \leq 10\%$,
and the light blue band refers to the corresponding  errors.
In the case of variation in $M_{\pi}$ by $-7\%$ in the S-wave, due to the difficulty in performing a Euclidean time
extrapolation at low energies, we estimate the error band from the spread in phase shifts versus
the number of time steps.}
\label{fig:phases_mpi_NNLO}
\end{figure}

\begin{figure}[htb]
\includegraphics[width=0.48\textwidth]{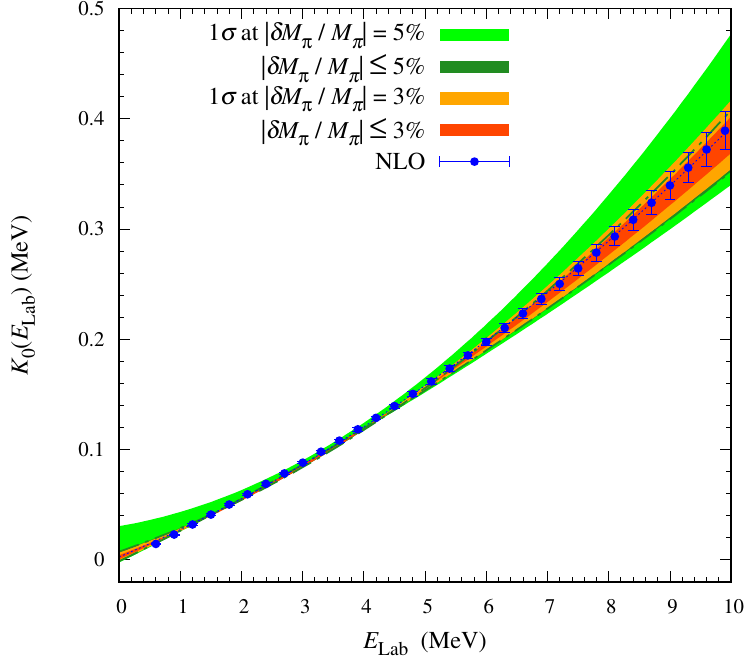}~~
\includegraphics[width=0.48\textwidth]{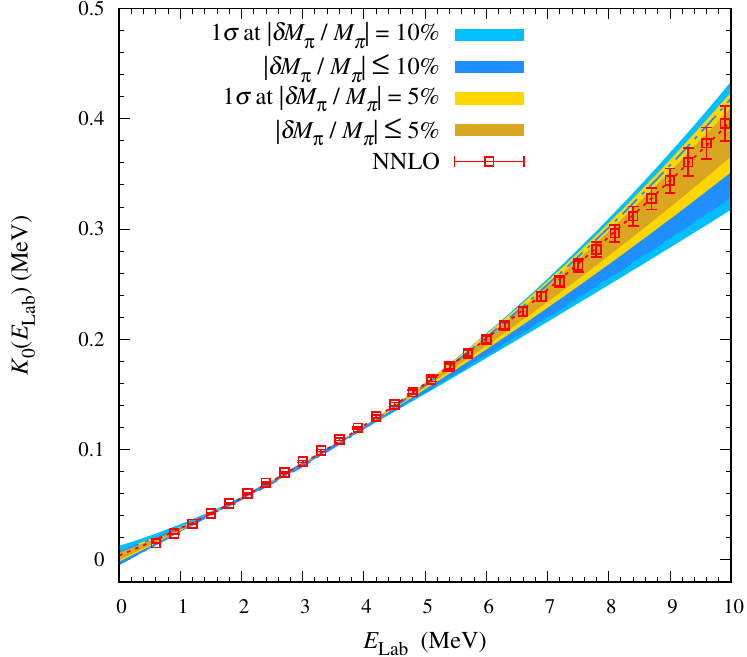}
\caption{Pion mass dependence of the S-wave effective range function. Left panel: NLO results. The blue
points refer to the results without pion mass variation, the red and dark green bands refer to changes
by $\pm 5\%$ and $\pm 10\%$, in order, and the orange and light green bands are the corresponding $1\sigma$
uncertainties. Right panel: NNLO results. The red squares refer to the results without pion mass variation,
the golden and dark blue bands refer to changes by $\pm 5\%$ and $\pm 10\%$, in order, and the yellow
and light blue bands are the corresponding $1\sigma$ uncertainties. 
}
\label{fig:K0}
\end{figure}

\begin{figure}[htb]
\includegraphics[width=0.48\textwidth]{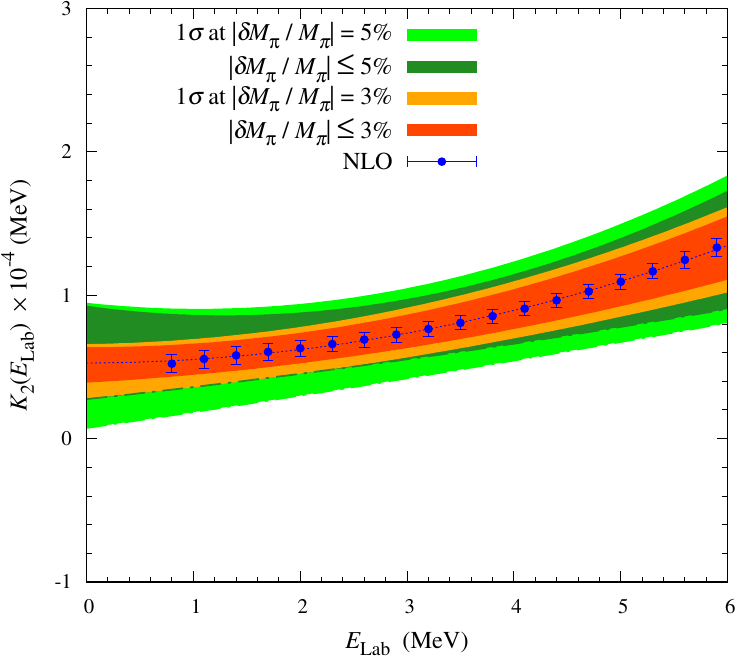}~~
\includegraphics[width=0.48\textwidth]{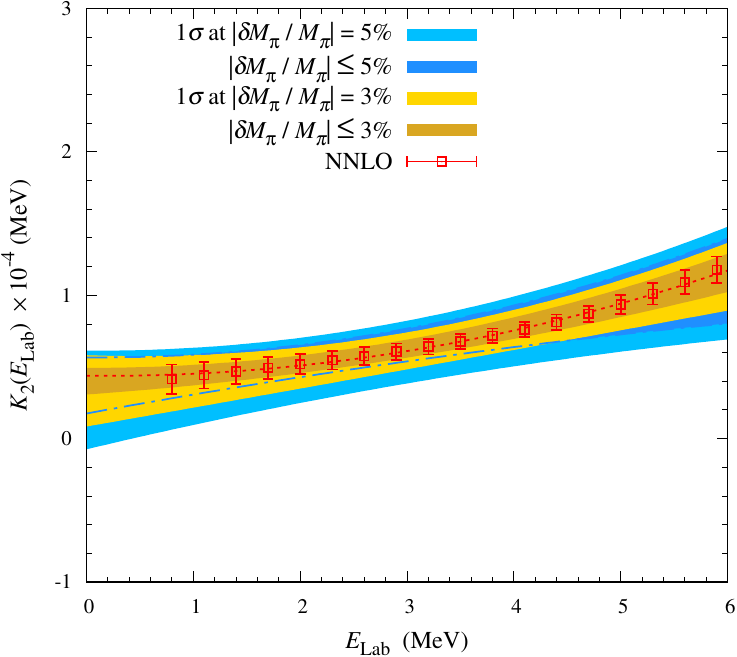}
\caption{Pion mass dependence of the D-wave effective range function. Left panel: NLO results. The blue
points refer to the results without pion mass variation, the red and dark green bands refer to changes
by $\pm 5\%$ and $\pm 10\%$, in order, and the orange and light green bands are the corresponding $1\sigma$
uncertainties. Right panel: NNLO results. The red squares refer to the results without pion mass variation,
the golden and dark blue bands refer to changes by $\pm 5\%$ and $\pm 10\%$, in order, and the yellow
and light blue bands are the corresponding $1\sigma$ uncertainties. 
}
\label{fig:K2}
\end{figure}

We now consider pion mass variations keeping $\alpha_{\rm EM} \simeq 1/137$ and $\theta\simeq 0$
fixed. In  Fig.~\ref{fig:phases_mpi_NLO}, we display the NLO results with variations of the pion
mass up to $\pm 3\%$ (inner red bands), together with the $1\sigma$ uncertainty of the 3\% variation
(outer orange bands) as well as the variations up to 5\% (inner dark green bands) and the
$1\sigma$ uncertainty of the 5\% variation (outer light green bands). As before, the left panel
gives the S-wave $\delta_0$ and the right panel the D-wave $\delta_2$ phase shift. The pertinent
$1\sigma$ uncertainties include all statistical and systematic errors properly propagated at this order.
Consider now in more detail the S-wave. For positive pion mass shifts, there is very little effect on
$\delta_0$, however, this is different for negative pion mass shifts. At around $\delta M_\pi/M_\pi
\simeq -5\%$, the additional repulsion unbinds the two-alpha system as seen by the phase shift starting
at zero.

In the D-wave, the effects of the pion mass variation are somewhat more pronounced, as
seen in the right panel of Fig.~\ref{fig:phases_mpi_NLO}. Here, the upper (lower) part
of the band refers to positive (negative) shifts in the pion mass. The
pion mass variation is also reflected in the
parameters of the D-wave resonance, which for a pion mass variation of $\pm 5\%$  are given by
\begin{eqnarray}
E_R &=& 2.57(6)~{\rm MeV}~~, ~~\Gamma_R = 1.22(21)~{\rm MeV}~\quad \delta M_\pi/M_\pi = +5\%~,\nonumber\\
E_R &=& 3.60(13)~{\rm MeV}~, ~~\Gamma_R = 3.56(89)~{\rm MeV}~\quad \delta M_\pi/M_\pi = -5\%~.
\end{eqnarray}

We now turn to the results at NNLO, showing the pertinent results for the S-wave
in the left panel of Fig.~\ref{fig:phases_mpi_NNLO} and for the D-wave in the right panel
of that figure. Consider first the S-wave, where we display results for pion mass variations in the
range $-7\% \leq \delta M_\pi/M_\pi \leq 10\%$. The critical value for $\delta M_\pi/M_\pi$, where the
two-alpha system becomes unbound, is moved to $-7\%$, where as positive changes of up to $10\%$ do not
lead to significant changes in the phase shift $\delta_0$. For the D-wave, we again find a larger
sensitivity (see right panel of  Fig.~\ref{fig:phases_mpi_NNLO}). This is again reflected in
the resonance parameters,
\begin{eqnarray}
E_R &=& 2.52(15)~{\rm MeV}~, ~~\Gamma_R = 0.92(33)~{\rm MeV}~\quad \delta M_\pi/M_\pi = +5\%~,\nonumber\\
E_R &=& 3.22(5)~{\rm MeV}~~, ~~\Gamma_R = 2.69(26)~{\rm MeV}~\quad \delta M_\pi/M_\pi = -5\%~.
\end{eqnarray}  
We note that both at NLO and NNLO, the variations of $E_R$ and $\Gamma_R$ are almost linear in the pion mass shift.

As noted, in our calculation at NNLO, the $^8$Be nucleus is slightly bound, which generates some of the
behaviour of the phase shifts close to zero energy. To overcome this, we also consider the pion mass
dependence of the S-wave effective range function $K_0(E_{\rm Lab})$ as well as the one of the
D-wave effective range function  $K_2(E_{\rm Lab})$, as defined in App.~\ref{app:modERE}.

Let us start with the S-wave.
In Fig.~\ref{fig:K0}, we show the pion mass variation of the S-wave effective range function with respect
to the results for our Universe at NLO (left panel) and NNLO (right panel). There appears to be little
effect on $K_0(E_{\rm Lab})$ at NLO, with a somewhat increased repulsion for negative pion mass shifts.
More precisely, there is some added repulsion for negative mass shifts.
This trend is also found at NNLO,  with some increase in strength.
We can quantify this by calculating the  shift in the
first parameter of the ERE, namely the inverse S-wave scattering length at NLO
\begin{equation}
  \frac{1}{a_0} = \left\{ \begin{matrix}
    -0.0017(12) & {\rm for}~~\delta M_\pi/M_\pi = -5\%~,\\
    -0.0025(3)  & {\rm for}~~\delta M_\pi/M_\pi = -3\%~,\\
    -0.0019(1)  & {\rm for}~~\delta M_\pi/M_\pi =  0~,~~~~\\
    -0.0016(1)  & {\rm for}~~\delta M_\pi/M_\pi = +3\%~,\\
    -0.0019(1)  & {\rm for}~~\delta M_\pi/M_\pi = +5\%~,\\
  \end{matrix} \right.
\end{equation}
and at NNLO
\begin{equation}
  \frac{1}{a_0} = \left\{ \begin{matrix}
    +0.0011(6)  & {\rm for}~~\delta M_\pi/M_\pi = -10\%~,\\
    -0.0016(1)  & {\rm for}~~\delta M_\pi/M_\pi = -5\%~,\\
    -0.0014(1)  & {\rm for}~~\delta M_\pi/M_\pi =  0~,~~~~\\
    -0.0013(1)  & {\rm for}~~\delta M_\pi/M_\pi = +5\%~,\\
    -0.0021(1)  & {\rm for}~~\delta M_\pi/M_\pi = +10\%~,\\
  \end{matrix} \right.
\end{equation}
all in units of MeV. We note that the shifts at NNLO are a bit larger than the ones at NLO,
which can be traced back to the fact that there  is more short-range repulsion in the NNLO interaction 
and thus it is less sensitive to the pion mass dependent corrections. Clearly, the NNLO calculation should be
considered more reliable.

Consider now the D-wave.
In Fig.~\ref{fig:K2}, we show the pion mass variation of the D-wave effective range function with respect
to the results for our Universe at NLO (left panel) and NNLO (right panel). The effect on 
$K_2(E_{\rm Lab})$ is quite pronounced, it is smallest where the phase shift passes through the resonance.
We can quantify this by calculating the  shifts in the inverse D-wave scattering length, at NLO
first parameter of the ERE, namely the inverse D-wave scattering length at NLO
\begin{equation}
  \frac{1}{a_2} = \left\{ \begin{matrix}
    9.30(2)  & {\rm for}~~\delta M_\pi/M_\pi = -5\%~,\\
    6.19(5)  & {\rm for}~~\delta M_\pi/M_\pi = -3\%~,\\
    5.27(5)  & {\rm for}~~\delta M_\pi/M_\pi =  0~,~~~~\\
    3.79(6)  & {\rm for}~~\delta M_\pi/M_\pi = +3\%~,\\
    2.42(12) & {\rm for}~~\delta M_\pi/M_\pi = +5\%~,\\
  \end{matrix} \right.
\end{equation}
and at NNLO
\begin{equation}
  \frac{1}{a_2} = \left\{ \begin{matrix}
    5.49(6)   & {\rm for}~~\delta M_\pi/M_\pi = -5\%~,\\
    4.95(8)   & {\rm for}~~\delta M_\pi/M_\pi = -3\%~,\\
    4.35(10)  & {\rm for}~~\delta M_\pi/M_\pi =  0~,~~~~\\
    3.02(4)   & {\rm for}~~\delta M_\pi/M_\pi = +3\%~,\\
    1.54(4)  & {\rm for}~~\delta M_\pi/M_\pi = +5\%~,\\
  \end{matrix} \right.
\end{equation}
all in units of $10^{-5}\,$MeV$^3$. Again, we find somewhat reduced changes at NNLO compared to NLO.

\ subsubsection{Alpha-alpha scattering with varying {$\alpha_{\rm EM}$}}
\label{sec:aEM}

\begin{figure}[htb]
\includegraphics[width=0.48\textwidth]{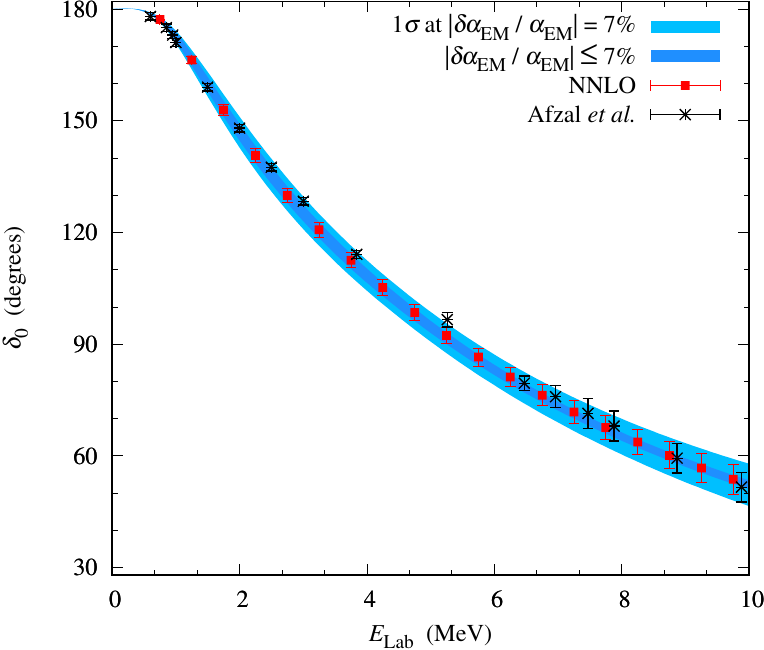}~~
\includegraphics[width=0.48\textwidth]{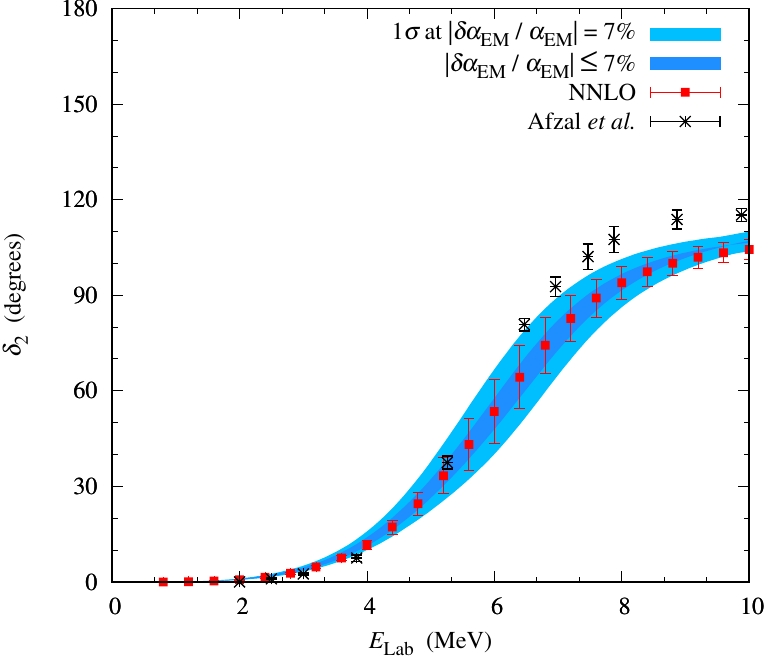}
\caption{Dependence of the $\alpha$-$\alpha$ phase shifts on the fine-structure constant
at NNLO. Left panel: S-wave  phase shift $\delta_0$ versus the energy in the laboratory system, $E_{\rm lab}$.
Right panel: D-wave  phase shift $\delta_0$ versus the energy in the laboratory system, $E_{\rm lab}$.
The black crosses refer to the experimental data, the red squares are the results for
$\delta \alpha_{\rm EM} = 0$, the dark blue band corresponds to variations in $ \alpha_{\rm EM}/\alpha_{\rm EM}
\leq 7\% (5\%)$, for the S-(D-)wave and the light blue band represents the corresponding $1\sigma$ error.}
\label{fig:phases_aEM_NNLO}
\end{figure}

Here, we consider the influence of variations in the fine-structure constant on the
$\alpha$-$\alpha$ phase shifts. Despite the various sources contributing to this
type of modifications as discussed in Sec.~\ref{sec:alpha-em-dependence}, we find
that the phase shifts are little affected by variations in $\alpha_{\rm EM}$, as
shown in Fig.~\ref{fig:phases_aEM_NNLO} for the NNLO results. Here,
variations of $\alpha_{\rm EM}$ up to $\pm 7\%$ are displayed, where the upper (lower)
part of the band refers to positive (negative) shifts in the fine-structure constant.
We see that the variation in $\alpha_{\rm EM}$ has essentially no effect on the phase
shifts. This can be explained as follows: By far the largest EM effect is the long-range
Coulomb interaction between the two clusters. Now we are measuring the phase shifts
with respect to the Coulomb-modified effective range expansion (see Appendix~\ref{app:modERE}),
and thus this dominant effect is already taken care of. In contrast to the bound state
energies (see Sec.~\ref{sec:energ}), the effect of the variation of the remaining, shorter-ranged
EM corrections appears to be insignificant.

\subsubsection{Remarks on the $\theta$-dependence of alpha-alpha scattering}
\label{sec:theta_res}

In Sec.~\ref{sec:theta}, we had shown that up to NLO, we can get the $\theta$-dependence of
the $\alpha$-$\alpha$ scattering phase shifts directly from the $\theta$-dependence of the
pion mass.  Therefore, we can directly translate the pion mass dependence of $\delta_{0,2}$
into a $\theta$-dependence. The depicted bands of the S-wave phase shifts for $\delta M_\pi/M_\pi =
-3\%$ and $-5\%$ in Fig.~\ref{fig:phases_mpi_NLO} correspond to a variation of $\theta = 0.7$ and
$0.9$, respectively. At such values of $\theta$, the di-proton and the di-neutron are bound and
element generation would proceed differently, for details see Ref.~\cite{Lee:2020tmi}.

We also note that a simultaneous variation of the light quark masses and $\theta$ can lead to a mutual (partial) compensation of effects, or to a mutual amplification. The latter case appears when $0 < |\theta| < \pi$ and at the same time  $\delta \hat{m}/\hat{m} < 0$ as both result in a decrease of the pion mass. If one the other hand has $0 < \delta \hat{m}/\hat{m} \leq 10\%$ one can always find a value for $\theta$ such that $M_\pi(\hat{m},\theta)=M_{\pi, \text{phys}}$ and nuclear physics would not be altered drastically (at least up to the order we are considering here).

\section{Summary and outlook}

In this work, we have considered the fundamental process of $\alpha$-$\alpha$ scattering based on
{\em ab initio} calculations in the framework of Nuclear Lattice Effective Field Theory, both
for the physical values of the light quark mass, the fine-structure constant, and the QCD
$\theta$-angle, as well as for variations in these parameters. The main findings of this work
can be summarized as follows:
\begin{itemize}
\item
  Due to improvements in the Adiabatic Projection Method compared to the pioneering study
  of $\alpha$-$\alpha$ scattering in Ref.~\cite{Elhatisari:2015iga}, we obtain a very good
  description of the S- and D-wave phase shifts up to energies $E_{\rm lab} \simeq 10\,$MeV
  at NNLO in the chiral expansion.
\item
  For the study of the variations under changes of the pion mass with $|\delta M_\pi/M_\pi| \leq 10\%$,
  we rely on the pion mass dependent nuclear Hamiltonian worked out in Ref.~\cite{Lahde:2019yvr}.
  To this orer, the $^8$Be nucleus is slightly bound.
  In the S-wave phase shift, we find a dramatic effect (unbinding of the two-alpha system) for
  changes of $-5\%$ and $-7\%$ at NLO and NNLO, respectively. We have also considered the pion mass
  variation of the S-wave effective range function, which is less sensitive to the binding issue
  and shows an added repulsion for negative pion mass shifts.  This additional repulsion will certainly
  impact the position and the lifetime of $^8$Be. The pion mass variation on the D-wave is somewhat
  more pronounced, as seen by the effect on the corresponding resonance parameters and also
  by the D-wave effective range function.
\item
  The dominant electromagnetic effect on the $\alpha$-$\alpha$ scattering phase shifts is the
  long-ranged Coulomb potential that is included exactly by using a spherical wall with
  Coulomb boundary conditions. Taking this effect into account via the Coulomb-modified ERE,
  we find very small effects of variations of $\alpha_{\rm EM}$ on the S- and D-wave phase shifts.
\item
  We have shown that up-to-and-including NLO in the chiral expansion, the dependence of the
  $\alpha$-$\alpha$ scattering phase shifts on the QCD $\theta$-angle is entirely given by the
  $\theta$-dependence of the pion mass.
\end{itemize}
In summary, we find that $\alpha$-$\alpha$ scattering (not unexpectedly) sets weaker constraints on the
variation of the light quark masses and the fine-structure constant than that given by the 
closeness of
the $3\alpha$ threshold to the Hoyle state. However, as discussed in detail e.g. in
Refs.~\cite{Oberhummer:2000zj,Lahde:2019yvr}, this requires stellar modelling which introduces
some model-dependence. In contrast to that, the investigation of $\alpha$-$\alpha$ scattering discussed
here is truly {\em ab initio} and not affected by such effects. Still, 
to further improve these calculations, a better determination of the pion mass dependence
of the singlet and triplet NN scattering lengths from lattice QCD is mandatory.

\section*{Acknowledgements}
We are grateful for discussions with members of the Nuclear Lattice Effective Field Theory Collaboration.
We gratefully acknowledge funding by  the Deutsche Forschungsgemeinschaft
(DFG, German Research Foundation) and the NSFC through the funds provided  to  the  Sino-German
Collaborative  Research  Center  TRR110  ``Symmetries  and  the  Emergence  of  Structure in  QCD''
(DFG  Project  ID 196253076  -  TRR  110,  NSFC Grant  No.  12070131001),
the Chinese Academy of Sciences (CAS) President's International Fellowship Initiative (PIFI)
(Grant No. 2018DM0034), Volkswagen Stiftung  (Grant  No.  93562),  the European Research Council (ERC) under the
European Union's Horizon 2020 research and innovation programme (grant agreement No. 101018170), the U.S. Department of Energy (DE-SC0013365 and DE-SC0021152) and the Nuclear Computational Low-Energy
Initiative (NUCLEI) SciDAC-4 project (DE-SC0018083) and the Scientific and Technological Research Council of Turkey (TUBITAK project no. 120F341).
The authors gratefully acknowledge the Gauss Centre for Supercomputing e.V. (www.gauss-centre.eu) for funding this project by providing computing time on the GCS Supercomputer JUWELS at Jülich Supercomputing Centre (JSC).
Further computational resources were provided by the
Oak Ridge Leadership Computing Facility through the INCITE award
``Ab-initio nuclear structure and nuclear reactions'',
and by the JSC on the JURECA-DC supercomputer.


\appendix
\section{Euclidean time extrapolation}
\label{app:Ete}

\begin{figure}[htb]
\includegraphics[width=0.49\textwidth]{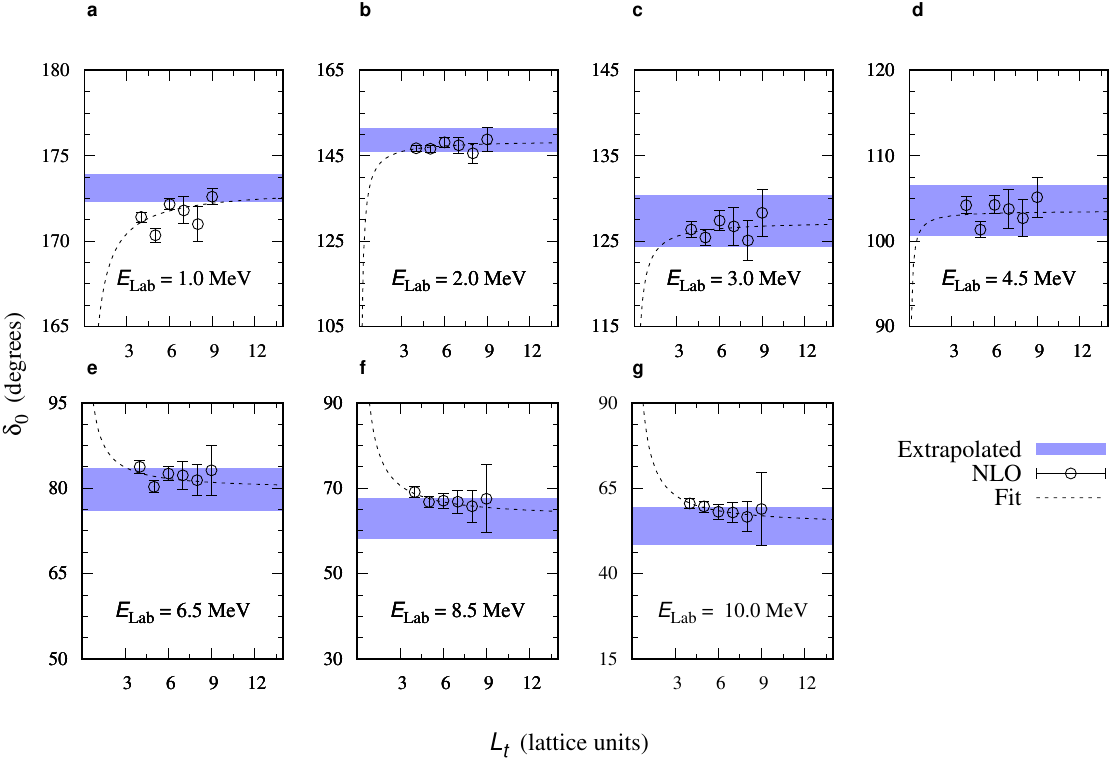}
\hfill\includegraphics[width=0.49\textwidth]{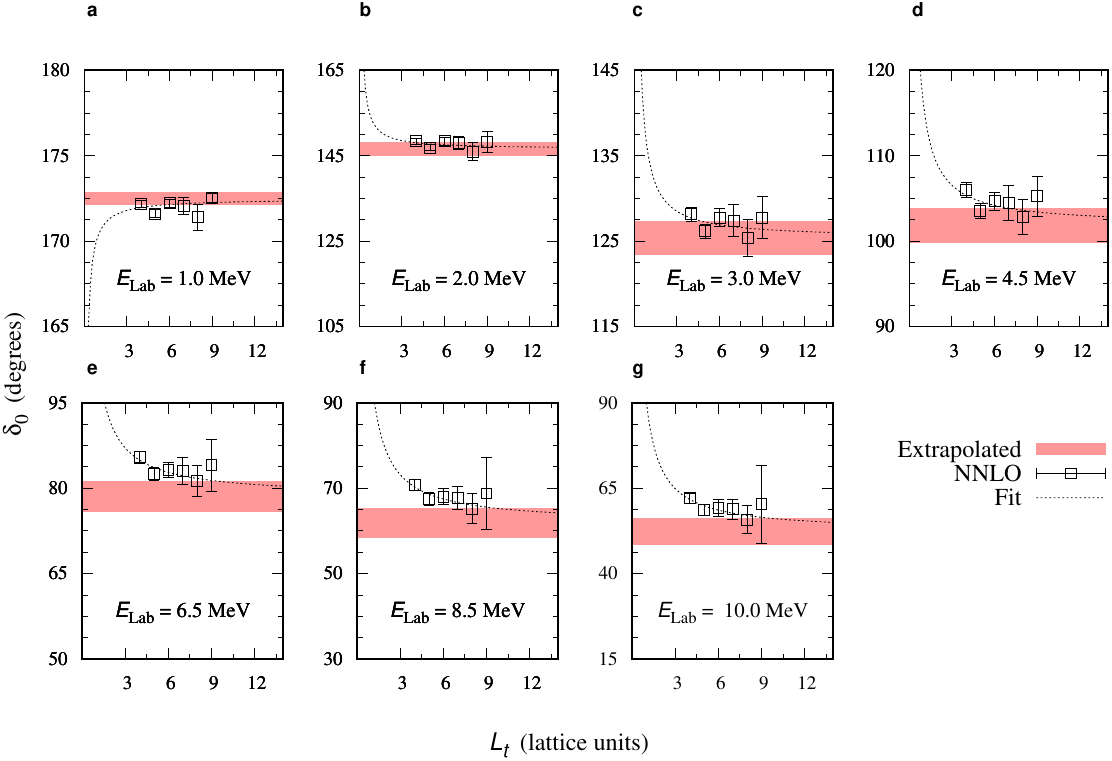}
\caption{NLO (left panel) and NNLO (right panel) results for the S-wave phase shift $\delta_0$ 
versus $L_t$ for the lab energies
$E_{\rm lab} =1.0\,$MeV, $2.0\,$MeV, $3.0\,$MeV, $4.5\,$MeV, $6.5\,$MeV, $8.5\,$MeV and
$10.0\,$MeV, respectively. The theoretical errors indicate the $1\sigma$ uncertainty due to the
MC errors. The dotted lines are fits to the data and used to extapolate to the $L_t\to\infty$
lomit. The hatched areas represent the $1\sigma$ error of the extrapolation.}
\label{fig:platS}
\end{figure}

\begin{figure}[htb]
\includegraphics[width=0.49\textwidth]{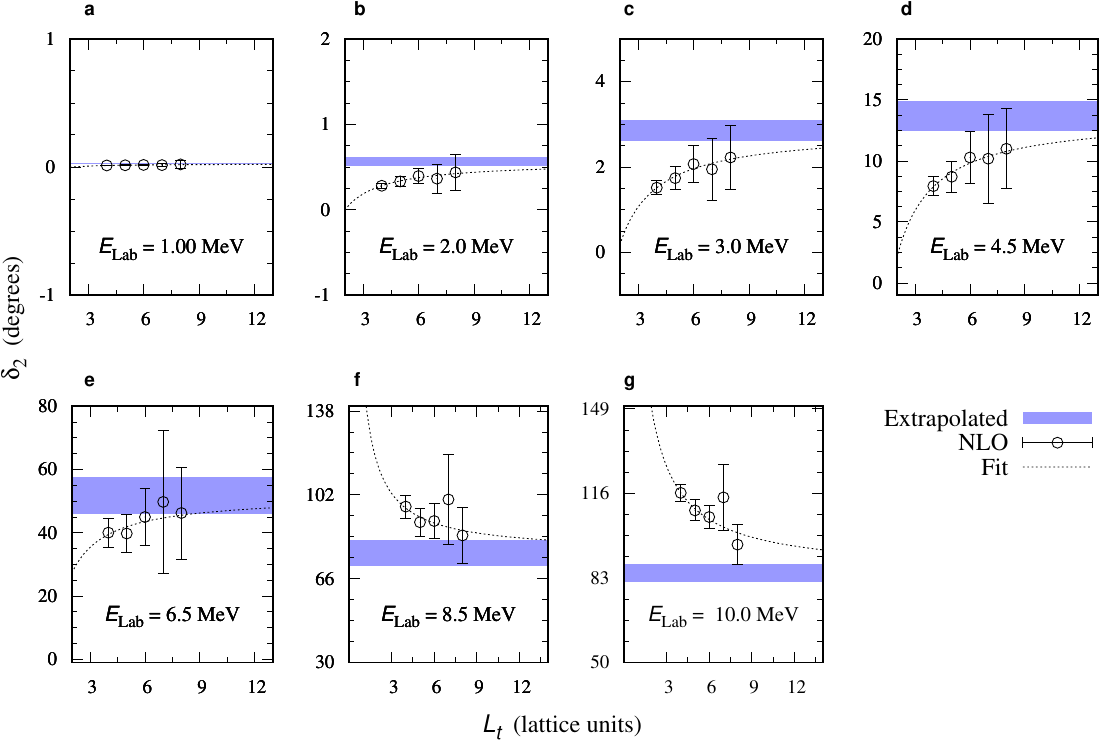}~~
\includegraphics[width=0.49\textwidth]{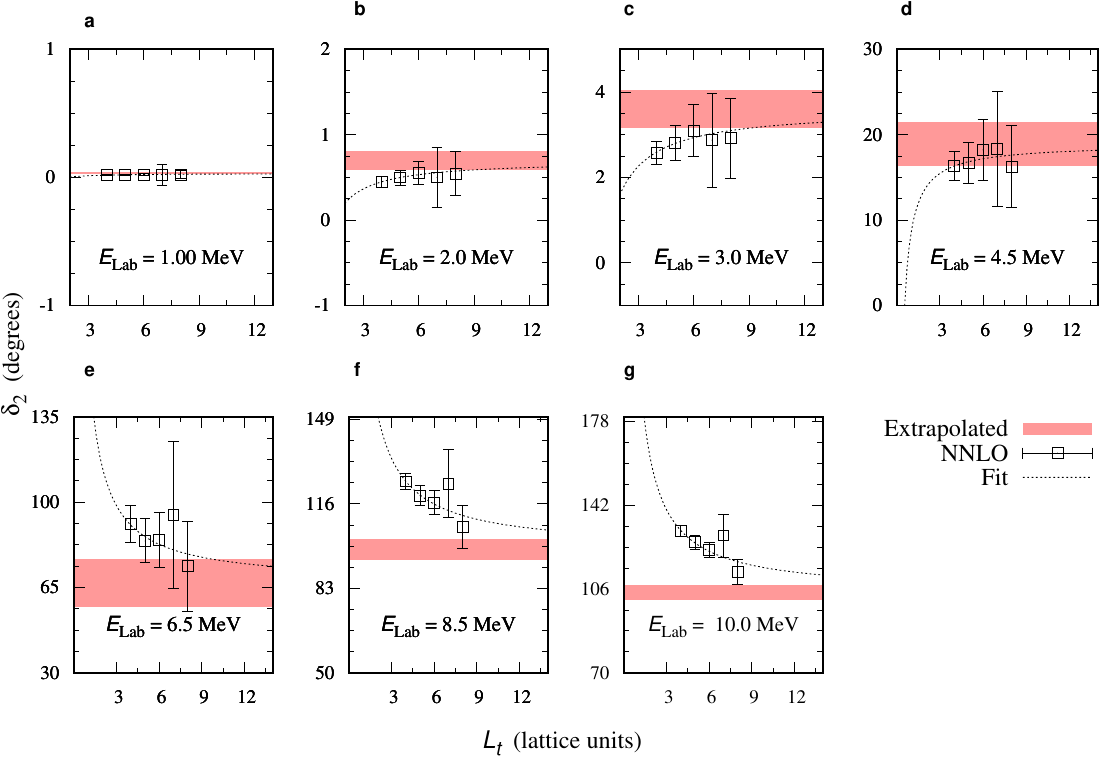}
\caption{NLO (left panel) and NNLO (right panel) results for the D-wave phase shift $\delta_2$ 
versus $L_t$ for the lab energies
$E_{\rm lab} =1.0\,$MeV, $2.0\,$MeV, $3.0\,$MeV, $4.5\,$MeV, $6.5\,$MeV, $8.5\,$MeV and
$10.0\,$MeV, respectively. The theoretical errors indicate the $1\sigma$ uncertainty due to the
MC errors. The dotted lines are fits to the data and used to extapolate to the $L_t\to\infty$
lomit. The hatched areas represent the $1\sigma$ error of the extrapolation.
}
\label{fig:platD}
\end{figure}

We perform the AFQMC simulations and construct the radial adiabatic transfer matrices for the S-wave and
D-wave channels from $L_t=4$ to $L_t =10$. Based on that, we compute the pertinent phase shifts
with errors calculated using a jackknife analysis of the MC data. In Figs.~\ref{fig:platS}
and \ref{fig:platD}
we show the NLO and NNLO results for the S- and D-wave phase shifts, respectively.

The dashed lines in these figures are the exponential curves used in the extrapolation
to the limit $L_t \to \infty$. This is achieved by including some residual dependence from an excited
state at an energy $\Delta E$ above the ground state, utilizing the ansatz:
\begin{equation}
\delta_{\ell} (L_t,E) = \delta_{\ell}(E) + c_{\ell}(E)\,\exp[-\Delta E_{\ell} \, L_t\, a_t]~,~~ {\ell}= 0,2~,
\end{equation}  
where the $c_{\ell}(E)$ and $\Delta E_{\ell}$ are fit parameters. 
As the gap between the $\alpha$-$\alpha$ threshold and these excited states is rather large,
one finds a fast convergence as exhibited in these figures. There, the hatched areas represent
the $1\sigma$ deviation errors of the extrapolations, including the propagated MC errors of the
data points.

\section{The Coulomb modified ERE}
\label{app:modERE}

Here, we collect the formulas for the Coulomb-modified ERE that was used above at
NLO and NNLO. The Coulomb modified ERE takes the
form~\cite{Bethe:1949yr,Jackson:1950zz,vanHaeringen:1981pb,Konig:2012prq}
\begin{equation}
K_\ell(p) = C^2_{\eta,\ell} p^{2\ell+1} \cot[\delta_\ell(p)] + \gamma h_\ell(p)
= -\frac{1}{a_\ell} + \frac{1}{2}r_\ell p^2  - \frac{1}{4}P_\ell p^4 + {\cal O}(p^6)~,  
\label{eq:modERE}
\end{equation}
for a partial wave with angular momentum $\ell$ and  $p$ is the relative momentum
of the two scattering clusters. $K_\ell(p)$ is also called the effective-range function for
angular momentum $\ell$. The factor $C^2_{\eta,\ell}$ is defined as
\begin{equation}
C^2_{\eta,\ell} = \frac{2^{2\ell}}{[(2\ell+1)!]^2} C^2_{\eta,0} \prod_{s=1}^\ell (s^2+\eta^2)~,
\end{equation}
where $C^2_{\eta,0}$ is the conventional Sommerfeld factor,
\begin{equation}
C^2_{\eta,0} = \frac{2\pi\eta}{e^{2\pi\eta}-1}~,
\end{equation}
with $\eta = \gamma / (2p)$. Here, $\gamma$ is the Coulomb parameter given by
\begin{equation}
  \gamma = 2 \mu\, \alpha_{\rm EM} \,Z_1 Z_2~,
\end{equation}  
where $\mu$ is the reduced mass of the two-alpha system and $Z_1=Z_2=2$ are the
charges of the two $\alpha$-particles.  Finally, the factor $h_\ell (p)$ in \eqref{eq:modERE}
is given by
\begin{equation}
h_\ell(p) = p^{2\ell} \frac{C^2_{\eta,\ell}}{C^2_{\eta,0}} \,\left( {\rm Re}[\psi(i\eta)] - \log|\eta|\right)~,
\end{equation}  
where $\psi(z)= \Gamma'(z)/\Gamma(z)$, in which the prime denotes differentiation.

\section{Bound state energies for varying pion masses}
\label{app:bsmpi}
Here, we collect the derivatives of the various ground state energies and the
energy of the Hoyle state with respect to the pion mass as a function of
the parameters $\bar{A}_s$ and $\bar{A}_t$, using the updated values for
$x_1$ and $x_2$ collected in Sec.~\ref{sec:Mpi-theta-dependence} (for details,
see Ref.~\cite{Epelbaum:2013wla}), 
\begin{align}
\left. 
\frac{\partial E_4^{}}{\partial M_\pi^{}} 
\right|_{M_\pi^\mathrm{ph}} = &
- 0.339(5) \, \bar A_s^{}
- 0.698(4) \, \bar A_t^{} 
+ 0.042(10)~,
\label{resultE_4} \\
\left. 
\frac{\partial E_8^{}}{\partial M_\pi^{}} 
\right|_{M_\pi^\mathrm{ph}} = &
- 0.796(31) \, \bar A_s^{}
- 1.584(22) \, \bar A_t^{} 
+ 0.098(25)~,
\label{resultE_8} \\
\left. 
\frac{\partial E_{12}^{}}{\partial M_\pi^{}} 
\right|_{M_\pi^\mathrm{ph}} = &
- 1.519(27) \, \bar A_s^{}
- 2.884(19) \, \bar A_t^{}
+ 0.174(46)~, 
\label{resultE_12} \\
\left. 
\frac{\partial E_{12}^\star}{\partial M_\pi^{}} 
\right|_{M_\pi^\mathrm{ph}} = &
- 1.589(12) \, \bar A_s^{}
- 3.025(9) \, \bar A_t^{} 
+ 0.194(47)~, 
\label{resultE_12s}
\end{align}
where the error in the parenthesis is the combined statistical one from the AFQMC calculation
and the systematic one due the uncertainties in $x_1$ and $x_2$.


\end{document}